\documentclass[preprint,prl,showpacs,preprintnumbers,amsmath,amssymb]{revtex4}

\usepackage{txfonts}    
\usepackage{bm}
\usepackage{color}
\definecolor{AAMgray}{gray}{0.45}
\usepackage[dvipdfmx]{graphicx} 
\newcommand{\jj }{\mathcal{J} }

\newcommand{\veps }{\epsilon} 
\newcommand{\eps }{\epsilon}

\usepackage{amsmath,amssymb}

\providecommand{\dd}{\mathrm{d}}
\providecommand{\e}{\mathrm{e}}
\providecommand{\Tr}{\mathrm{Tr}}

\providecommand{\bm}{\boldsymbol} 
\usepackage{braket}
\usepackage{mathrsfs} 
\newcommand{\ketbra}[2]{\ket{#1}\!\bra{#2}}

\definecolor{AAMgray}{gray}{0.45}

\newcommand{\AAMdoi}{10.1088/1361-6587/ae86de}

\newcommand{\AAMtitleNotice}{%
  \par\vspace{4pt}%
  {\normalfont\footnotesize\color{AAMgray}%
  \noindent
  \begin{minipage}{0.96\textwidth}
  \large \ \\This is the author accepted manuscript (AAM/AM), published in Plasma Physics and Controlled Fusion, Vol.68 (2026).\\[1mm]
  DOI: \AAMdoi
  \end{minipage}%
  \par}%
}

\makeatletter
\let\AAM@orig@pacs@produce\@pacs@produce
\def\@pacs@produce#1{%
  \AAM@orig@pacs@produce{#1}%
  \AAMtitleNotice
}
\makeatother

\begin{document}

\title{Out-of-time-ordered correlators for turbulent fields:\\ a quantum-classical correspondence}

\author{Motoki Nakata$^{1,2}$}
\affiliation{$^1$Faculty of Arts and Sciences, Komazawa University, Tokyo 154-8525, Japan\\ 
$^2$RIKEN Center for Interdisciplinary Theoretical and Mathematical Sciences (iTHEMS), Saitama 351-0198, Japan}

%
%


\begin{abstract}
An extended formulation of out-of-time-ordered correlators (OTOCs), which quantify noncommutative operator growth and information scrambling 
in quantum many-body systems, is developed for turbulence dynamics as a representative of non-canonical Hamiltonian systems.  
Based on the Wigner-Weyl transform and the Moyal bracket formalism, the semiclassical limit of OTOC for turbulent plasmas governed 
by the Hasegawa-Mima equation is derived as an ensemble-averaged squared Lie-Poisson bracket between two chosen functionals of the turbulent fields. 
The classical-limit OTOC provides a quantitative measure of how a variational perturbation applied to one functional propagates across scales
in the turbulent dynamics and how it affects another functional at a later time, thereby capturing scale-dependent or field-dependent transfer processes. 
In a quasilinear approximation with a strong zonal flow, we provide a closed analytic expression 
of the classical-limit OTOC to characterize the interaction between zonal and non-zonal modes.   
An asymptotic analysis shows that the OTOC grows quadratically at early time, while in the long-time strong-shear regime it approaches 
a finite saturated value with an inverse-square algebraic dependence. 
This behavior is attributed to zonal-flow shearing, which rapidly scrambles the non-zonal perturbation toward higher wavenumbers, 
thereby reducing the low-wavenumber non-zonal content that can feed back onto large-scale zonal modes.
\end{abstract}

\pacs{XXXX}

\maketitle
\section{Introduction}
Turbulence is a prototypical nonlinear dynamical system that exhibits sensitivity to perturbations over a wide range of interacting scales, 
and is one of the most important issues in plasma and fluid physics\cite{Horton}. 
Such sensitivity is often discussed in the framework of classical chaos theory and quantified by Lyapunov analysis and its extended variants. 
Notable recent progress in the neutral fluid context has been made by combining covariant Lyapunov vectors with data assimilation techniques 
in the tangent-space dynamics of perturbations\cite{Inubushi1, Inubushi2}.

Related diagnostics have also been developed in terms of classical decorrelators, 
which quantify the separation between nearby trajectories or the spreading of a localized perturbation. 
Such quantities have been used to characterize butterfly effects and perturbation spreading 
in many-body classical systems and fluids\cite{Das2018,Bilitewski2018,Murugan2021}.

There is also growing interest in introducing information-theoretic diagnostics into turbulence. 
Tanogami and Araki recently proposed statistical information measures, e.g., mutual-information-based or information-thermodynamic quantities, 
to quantify scale-to-scale information transfer in turbulent cascades and to derive inequality constraints on such transfer\cite{Tanogami1,Tanogami2}. 
These approaches revealed statistical aspects of information flow, complementing wavenumber spectra and correlation functions. 
In addition, a quantum-information-inspired modal analysis for multi-field plasma turbulence is proposed\cite{Yatomi1, Yatomi2}, 
where the von Neumann entropy clarified a nontrivial transition between coherent (dominated by zonal flows) and incoherent (dominated by vortices) 
turbulence states. 

Despite these significant advances, a challenge remains beyond Lyapunov analysis which evaluates global instabilities of the system, 
and information-theoretic measures which primarily capture statistical correlations. 
In order to directly analyze the dynamical causality between specific modes or fields, e.g., large-scale zonal flows, drift-wave vortices, 
density, temperature, and magnetic fields, an observable-dependent measure that preserves the conservation properties in the underlying Poisson structure 
is required.
To this end, the out-of-time-ordered correlators (OTOCs) provide a unique and powerful perspective. 

In quantum many-body systems, the chaotic behavior cannot be captured by exponential separation of nearby trajectories 
because of the ill-posedness of the deterministic trajectory. 
Instead, the corresponding argument of sensitivity has been developed from a different viewpoint, i.e., 
noncommutative operator growth and information scrambling quantified by the OTOC. 
The concept of OTOC has originally been introduced in the context of superconductivity\cite{Larkin1969}, 
and revitalized in modern studies of quantum chaos\cite{Hashimoto2017,Rozenbaum2017,TsujiShitaraUeda2018,XuSwingle2024,Fujii2025}, 
black hole thermodynamics/quantum gravity\cite{Kitaev2014,Shenker2014,Maldacena2016,Swingle2018}, 
and quantum computing\cite{Yoshida2019, Mi2021,Google2025}.
As will be shown in Sec. 2, the OTOC is defined by the squared commutator of the time-evolving Heisenberg operators 
and measures how initially local operators spread over many degrees of freedom, thereby making local information effectively delocalized.
This is physically interpreted as nonlocal operator spreading and scrambling in real space or in the space of degrees of freedom, i.e., 
the ``butterfly effect'' for quantum systems in large degrees of freedom. 

These developments raise a natural question in classical turbulence of plasmas and fluids, i.e., 
whether an OTOC-like quantity can be formulated so as to quantify how a perturbation applied to a chosen functional of turbulent fields
is transferred across turbulence dynamics and feeds back to another functional. 
However, a subtle point is that the naive classical limit of the quantum OTOC trivially vanishes as $\hbar \to 0$. 
Nontrivial information is then obtained by extracting the leading semiclassical contribution from the quantum OTOC\cite{Jalabert2018,Michel2025}. 
Another difficulty lies in the fact that, since the turbulence dynamics is described as a non-canonical Hamiltonian formalism, 
the canonical OTOC cannot simply be applied. 

The above questions are also related to classical decorrelators, but the present study emphasizes a different aspect. 
Classical decorrelators are powerful diagnostics of trajectory separation or local perturbation spreading in canonical many-body systems. 
However, trajectory-distance-based measures for turbulent fields complicate non-unique choices of metric, projection, field component, and perturbation protocol, 
making non-local or projected perturbation-response relations difficult to diagnose. 
It is also desirable that admissible perturbations respect the underlying geometric constraints, e.g., Casimir invariants, in non-canonical Hamiltonian systems. 
The OTOC considered here can address these issues by defining the response directly through the non-canonical Lie-Poisson bracket between physically selected functionals. 

In this work, we develop an extended formulation of OTOCs for classical turbulence dynamics in the context of non-canonical Hamiltonian systems.
Focusing on the wave-vortex plasma turbulence governed by the Hasegawa-Mima (H-M) equation\cite{HasegawaMima1977}, we derive the classical-limit OTOC as
the ensemble-averaged squared Lie-Poisson bracket between two chosen functionals of the turbulent fields. 
A key advantage is that the relevant degrees of freedom, scales, and fields can be selected through the choice of functionals and projection operators. 
This enables a scale-dependent or field-dependent characterization of perturbation transfer processes, complementing the conventional turbulence diagnostics. 
We then apply this formulation to the interaction analysis between fine-scale non-zonal fluctuations and large-scale zonal flows, 
and the asymptotic scaling of the classical-limit OTOC is revealed.   

The rest of the paper is organized as follows. 
Section 2 introduces quantum OTOCs in the density-operator formulation, and derives the nontrivial semiclassical limit 
in terms of Wigner-Weyl transform and Moyal bracket expansion, including a natural extension to the non-canonical Poisson structure. 
The classical-limit OTOC is then applied to the H-M turbulence in Sec. 3, 
and the asymptotic time-lag scaling of the OTOC for the zonal-mode response to the non-zonal perturbations 
is discussed to highlight the physical meaning and its usefulness. 
Section 4 summarizes the results and discusses implications and future applications.
Also, some fundamental derivations are presented in Appendices. 
%

\section{Theory of out-of-time-ordered correlators: OTOCs}
In this section, we briefly introduce out-of-time-ordered correlators (or out-of-time order correlators in some literature), i.e., OTOCs, 
which have been developed for quantum many-body systems. 
OTOCs quantify chaotic behaviour of how rapidly local perturbations become non-localised or spread, 
by utilizing the growth of the commutator $[\hat{A}(t),\hat{B}(t_0)]$ between the time-evolving operator $\hat{A}$ at time $t$
and the initial operator $\hat{B}$ at time $t_0$. 
Then we construct a semiclassical limit of OTOC and explain its physical significance, 
which provides us with new insights to describe turbulence dynamics in plasmas and fluids. 

\subsection{OTOCs in quantum systems}
Let $\hat\rho(t)$ be a density operator describing either a pure or mixed state, 
i.e., $\hat{\rho}(t)= \ketbra{\psi}{\psi}$ for the pure state and $\hat{\rho}(t)= \sum_{j}p_{j}\ketbra{\psi_j}{\psi_j}$ 
for the mixed state with a non-negative weight (probability) $p_j \ge 0$ and $\sum_{j}p_j = 1$. 
Then $\Tr(\hat{\rho}(t))=1$ for both cases. 
The time evolution under a Hamiltonian operator $\hat{H}$ is governed by the von Neumann equation
\begin{equation}
i\hbar\,\partial_t\, \hat{\rho}(t) = \left [ \hat{H},\hat{\rho}(t) \right ],
\end{equation}
where $\hbar$ denotes the reduced Planck constant and the Lie bracket or commutator is defined 
as $[\hat{X},\hat{Y}]:= \hat{X}\hat{Y}-\hat{Y}\hat{X}$.
Although Eq. (1) is written in the Schr\"{o}dinger picture, 
we discuss the OTOC in the Heisenberg picture because the two descriptions are equivalent under the same unitary evolution.
In the Heisenberg picture, a physical quantity (or an observable) as an operator $\hat A(t)$ satisfies
\begin{eqnarray}
i\hbar\,d_t\, \hat{A}(t) & \!  =  \! & \left [ \hat{A}(t),\hat{H} \right ] + i\hbar\partial_t \hat{A},\\
\hat A(t) & \!   =   \! & \hat{U}^{\dagger}\,\hat{A}(t_0)\,\hat{U},\\ 
\hat U(t,t_0) & \!   =  \! & \e^{-i\hbar^{-1}(t-t_0)\,\hat{H}},
\end{eqnarray}
where the expectation value is given by $\langle \hat{A}(t) \rangle:=\Tr(\hat{\rho}(t_0)\,\hat{A}(t))$.
The dagger symbol denotes the Hermitian conjugate.
The explicit time dependence in $\hat{A}$ and $\hat{H}$ is ignored in the following, 
i.e., $\partial_t \hat{A} =0$ and $\partial_t \hat{H} =0$. 
For time-dependent Hamiltonian, $\hat{U}(t,t_0)$ should be understood as a time-ordered exponential.

A quantum OTOC is defined by the state-averaged squared commutator (see References in Sec. 1): 
\begin{equation}
\mathcal{C}^{\rm q}_{\hat{A}\hat{B}}(t,t_0)
:=
\Tr\left(\hat{\rho}(t_0)\,\left [\hat{A}(t),\hat{B}(t_0) \right ]^\dagger \left [\hat{A}(t),\hat{B}(t_0) \right ]\right) \ge 0.
\end{equation}
The OTOC has been derived as a powerful theoretical framework 
to quantify the noncommutativity of physical operators referred as operator ``scrambling'', 
i.e., the process by which initially localized information of the state at $t_0$ becomes inaccessible to local measurements at $t$ 
due to operator growth with complex correlations.

The name ``out-of-time-ordered'' stems from the fact that Eq. (5) involves a sequence of operations 
that cannot be arranged in a simple temporal order, constructing a time-loop of $t_0  \rightarrow t \rightarrow t_0$. 
The commutator $[\hat{A}(t), \hat{B}(t_0)]$, where the first term is given by $\hat{U}^{\dagger}(t,t_0)\hat{A}(t_0)\hat{U}(t,t_0)\hat{B}(t_0)$, 
measures how an initial perturbation $\hat{B}$ at arbitrary $t_0$ interferes 
with a later observation $\hat{A}$ at $t$, by ``rewinding'' the system from $t$ back to $t_0$ after the action of $\hat{A}$. 
When $\hat{A}(t)$ remains substantially localized or its operator growth is negligible, 
the influence of the perturbation $\hat{B}(t_0)$ is minimal and the commutator also remains small.
However, once $\hat{A}(t)$ becomes highly nonlocal through interacting dynamics among the quantum elements, the action of $\hat{B}(t_0)$ 
and the subsequent evolution does not commute so that the squared commutator also grows. 
The growth of the OTOC, thus, provides a dynamical diagnostic of the operator noncommutativity, 
characterizing the nonlocal spreading speed of initially local perturbations. 
%
%
While such a sensitivity is conventionally analyzed by Lyapunov exponents in classical dynamics, 
the OTOC framework has extensively been applied to quantum computing and information dynamics in black hole physics. 
%
%

\subsection{Semiclassical limit of OTOC in Weyl-Wigner-Moyal formalism}
In this subsection, we connect the quantum OTOC in Eq. (5) to classical field dynamics. 
Following several earlier works\cite{Jalabert2018,Michel2025}, 
the derivation of a classical-limit OTOC is demonstrated based on the Weyl-Wigner-Moyal formalism for quantum-classical correspondence\cite{Weyl1927,Wigner1932,Moyal1949}.
For definiteness, we first consider a canonical phase space with coordinates $\bm{z}:=(\bm{q},\bm{p})\in\mathbb{R}^{2n}$ and symplectic matrix $\bm{J}$, 
where the extension to non-canonical system will be discussed later.   

Let $\mathcal{W}$ be a Weyl transform, which is also referred to as Weyl quantization or deformation quantization. 
A phase-space function (or more mathematically symbol) $F(\bm{z})$ is then mapped into a corresponding noncommutative operator $\hat F$:  
\begin{equation}
\hat{F} = \mathcal{W}[F(\bm{z})] := \frac{1}{(2\pi)^{2n}} \! \int \! d\bm{\xi} d\bm{\eta} \  \mathscr{F}\left [F(\bm{z}) \right] 
\exp \left [ i\left( \bm{\xi} \cdot \bm{\hat{q}} + \bm{\eta}\cdot \bm{\hat{p}} \right) \right ], 
\end{equation}
where $\mathscr{F}$ is Fourier transform on the phase space, i.e., $\mathscr{F}[F(\bm{z})] := \int \! d\bm{p}d\bm{q}\, F(\bm{q}, \bm{p}) 
\exp[-i(\bm{\xi}\! \cdot\! \bm{q} + \bm{\eta}\! \cdot\! \bm{p})]$. 
Note also that $[\bm{\hat{q}},\bm{\hat{p}}] =i\hbar\bm{I}_n$, and $\bm{I}_n$ means the $n \times n$ unit matrix. 
The inverse transform $\mathcal{W}^{-1}$ is called Wigner (or Wigner-Weyl) transform to map an operator $\hat{F}$ 
into a phase-space function $F(\bm{z})$: 
\begin{eqnarray}
F(\bm{z}) = \mathcal{W}^{-1}[\hat{F}] \!  & =  \! & \! \int \! d\bm{s} \exp \left ( -i\hbar^{-1}\bm{s}\cdot\bm{p} \right) 
\Braket{\bm{q}+\frac{\bm{s}}{2}|\hat{F}|\bm{q}-\frac{\bm{s}}{2}} \\ 
\! \!  & =  \! \! & \Tr \left (\hat{F}\hat{\Omega}(\bm{z}) \right ), 
\end{eqnarray}
where 
$\hat{\Omega}(\bm{z}):= \int \! d\bm{s} \exp(-i\hbar^{-1}\bm{s} \cdot  \bm{p})\,\Ket{\bm{q} \! -\! \bm{s}/2} \! \Bra{\bm{q} \! +\! \bm{s}/2}$. 
One can show their consistency by substituting Eq. (6) to Eq. (7), using Baker-Campbell-Hausdorff (BCH) formula. 

As shown in Eqs. (2) and (5), the operator evolution and the OTOC involve the multiplication of operators through the commutator. 
The Wigner transform of the commutator is given by 
\begin{eqnarray}
\mathcal{W}^{-1} \left [ (i\hbar)^{-1}[\hat{A},\hat{B}] \right ] 
 \!  & =  \!  & (i\hbar)^{-1}\left [ A(\bm{z})\star B(\bm{z})-B(\bm{z})\star A(\bm{z})\right ] \\
 \!  & =:  \!  & \{A(\bm{z}),B(\bm{z})\}_{\rm M}, 
\end{eqnarray}
where $\{\cdot,\cdot\}_{\rm M}$ denotes the Moyal bracket with so-called star product $\star$ defined as 
\begin{eqnarray}
A(\bm{z})\star B(\bm{z})
 \! \! & :=  \! \! & A(\bm{z})\exp \left[\frac{i\hbar}{2}\Big(\overleftarrow{\partial}_{\bm{z}}\cdot \bm{J}\cdot \overrightarrow{\partial}_{\bm{z}}\Big)\right]B(\bm{z}), \\
\bm{J}  \! \! & =  \! \! & \begin{pmatrix}
  0 & \bm{I}_n \\
  -\bm{I}_n & 0
\end{pmatrix}. 
\end{eqnarray}
Here, $\overleftarrow{\partial}_{\bm{z}}$ or $\overrightarrow{\partial}_{\bm{z}}$ 
acts on the function in the left side ($A$) or the right side ($B$). 
Applying Eq. (9) to the von Neumann equation in Eq. (1) yields the Wigner-Moyal equation
\begin{equation}
\partial_t W(t,\bm{z}) = \left \{H(\bm{z}),W(t,\bm{z}) \right \}_{\rm M},
\end{equation}
where $W(t,\bm{z}):= \mathcal{W}^{-1}[\hat{\rho}(t)]$ is so-called Wigner function on the phase space. 
Note that while the Wigner-Moyal equation is represented by the canonical phase-space coordinate $\bm{z} = (\bm{q}, \bm{p})$, 
all the quantum effects are described by the higher-order terms in $\hbar$ in the Moyal bracket.  
In recent years, the Wigner-Moyal equation 
has been actively applied as a wavekinetic theory that provides a reduced description of dispersive wave propagation 
and turbulence dynamics in plasmas\cite{Ruiz2016, Zhu2018, ZhuDodin2021, Dodin2024}.

Expanding Eq. (10) in $\hbar$ gives
\begin{eqnarray}
\{A(\bm{z}),B(\bm{z})\}_{\rm M}  \! \! & =  \! \! & \{A(\bm{z}),B(\bm{z})\}_{\rm PB} \nonumber \\
\qquad \qquad & - & \! \! \! \frac{\hbar^2}{24}A(\bm{z}) 
\Big(\overleftarrow{\partial}_{\bm{z}}\cdot \bm{J}\cdot \overrightarrow{\partial}_{\bm{z}}\Big)^3 B(\bm{z})
 +  \mathcal{O}(\hbar^4),
\qquad
\end{eqnarray}
where $\{A,B\}_{\rm PB}:=\partial_{\bm{q}} A \cdot \partial_{\bm{p}} B - \partial_{\bm{q}} B \cdot \partial_{\bm{p}} A$.  
The leading-order classical limit of $\hbar \! \rightarrow \! 0$ in the Moyal bracket is, thus, the canonical Poisson bracket $\{\cdot,\cdot\}_{\rm PB}$.

By utilizing the trace identity of
\begin{eqnarray} 
\Tr(\hat{X}\hat{Y})  \! & =  \! & \! \int \! \frac{d\bm{z}}{(2\pi\hbar)^{n}}\, 
\mathcal{W}^{-1}\left [\hat{X} \right ](\bm{z}) \star \mathcal{W}^{-1}\left [\hat{Y} \right ](\bm{z})  \nonumber \\
 \! & =  \! &  \! \int \! \frac{d\bm{z}}{(2\pi\hbar)^{n}}\, X(\bm{z})Y(\bm{z}), 
\end{eqnarray}
the quantum OTOC in Eq. (5) is expressed as 
\begin{eqnarray}
\mathcal{C}^{\rm q}_{\hat{A}\hat{B}}(t,t_0)
& = & \int \! \frac{d\bm{z}}{(2\pi\hbar)^{n}}\, W(t_0,\bm{z})\left ( \mathcal{W}^{-1}\left [ \left [\hat{A}(t),\hat{B}(t_0) \right ]^\dagger \right ] \star \mathcal{W}^{-1} \left [ \left [\hat{A}(t),\hat{B}(t_0) \right ] \right ] \right ) \nonumber \\
& = & \hbar^2 \int \! \frac{d\bm{z}}{(2\pi\hbar)^{n}}\, W(t_0,\bm{z})\, \Big ( \left \{A(t,\bm{z}),B(t_0,\bm{z}) \right \}_{\rm M}^{\dagger} \star \left \{A(t,\bm{z}),B(t_0,\bm{z}) \right \}_{\rm M} \Big ) \nonumber \\ 
& = & \hbar^2 \int \! \frac{d\bm{z}}{(2\pi\hbar)^{n}}\, W(t_0,\bm{z}) \left ( \big | \left \{A(t,\bm{z}),B(t_0,\bm{z}) \right \}_{\rm M} \big |^2 + \mathcal{O}(\hbar^2) \right ), \nonumber \\
& = & \hbar^2 \int \! d\bm{z}\, f_{\hbar}(t_0,\bm{z}) \left ( \big | \left \{A(t,\bm{z}),B(t_0,\bm{z}) \right \}_{\rm M} \big |^2 + \mathcal{O}(\hbar^2) \right ), 
\end{eqnarray}
where the boundary terms are assumed to vanish in the identity Eq. (15) 
when the integrand rapidly decreases towards the boundary, or the periodic condition is imposed. 
In the last equality of Eq. (16), the normalized distribution $f_{\hbar}$ is introduced, i.e., 
$f_{\hbar}(t,\bm{z}):= (2\pi\hbar)^{-n}\, W(t,\bm{z})$ with $\int d\bm{z}\, f_{\hbar}(t,\bm{z}) = 1$ because of 
the identity $\Tr(\hat{\rho}(t)) = 1$. 
Therefore, we found that the semiclassical counterpart is obtained by expanding the Moyal bracket in Eq. (16) 
up to the lowest non-vanishing order in $\hbar$. 

Since the Moyal bracket reduces to the Poisson bracket as $\{A, B\}_{\rm M} \to \{A, B\}_{\rm PB}$ in the lowest order, 
the quantum OTOC scales as $C^{q}_{\hat{A}\hat{B}} \! \sim \! \hbar^2  \{A(t), B\}_{\rm PB}^2$, thereby vanishing for taking simply $\hbar \! \rightarrow \!0$. 
This trivially means the vanishing noncommutativity of the operators in the lowest order for the classical systems. 
However, by isolating the semiclassical dynamical part from $C^{q}_{\hat{A}\hat{B}}$ in the leading-order quantum scaling factor of $\hbar^2$, 
we arrive at a straightforward definition of the nontrivial classical-limit OTOC as follows: 
\begin{eqnarray}
C^{\rm cl}_{AB}(t,t_0)  \! \! & :=  \! \! & \lim_{\hbar \rightarrow 0} \hbar^{-2}\mathcal{C}^{\rm q}_{\hat{A}\hat{B}}(t,t_0) \nonumber \\
 \! \! & = \! \! &  \left\langle \{A(t,\bm{z}),B(t_0,\bm{z})\}_{\rm PB}^2 \right\rangle_{\rm ens},
\end{eqnarray}
where the average operator $\langle X(\bm{z}) \rangle = \! \int \! d\bm{z} f_{\rm cl}(t_0, \bm{z}) X(\bm{z})$ 
is expressed with the classical phase-space distribution of $\lim_{\hbar  \rightarrow  0}f_{\hbar} = f_{\rm cl}$, 
and is evaluated as the ensemble average at the initial state, i.e., $\langle X \rangle = \langle X \rangle_{\rm ens}$. 
For numerical evaluations, this average can be estimated by sampling initial conditions $\bm{z}_0^{(n)}:=\bm{z}^{(n)}(t_0)$ from $f_{\rm cl}$, i.e., 
$ \langle X\rangle_{\rm ens} = N^{-1}\sum_{n=1}^{N}X[\bm{z}_0^{(n)}]$ for large $N$. 
The classical-limit OTOC in Eq. (17) characterizes how an infinitesimal perturbation applied to the observable $B$ at the initial time $t_0$ 
propagates, and emerges as a response in another observable $A$ at a later time $t$.
For canonical Hamiltonian systems, the semiclassical reduction of the quantum OTOC to the square of a classical Poisson bracket, 
as well as its connection to Lyapunov growth and saturation, has extensively been studied\cite{Jalabert2018,Michel2025}.
The derivation below follows this canonical quantum-classical correspondence at the leading semiclassical order, 
and then uses it as a starting point for the non-canonical generalization. 

It is emphasized that the classical-limit OTOC possesses a nontrivial temporal structure 
that reflects the out-of-time-ordered nature of its quantum counterpart with noncommutativity. 
While the squared Poisson bracket $\{A(t,\bm{z}), B(t_0, \bm{z})\}^2_{\rm PB}$ may superficially appear as a simultaneous correlation of $A$ and $B$, 
its evaluation inherently involves the ``pull-back'' of information from an evolved state $A(t, \bm{z})$ to the initial phase-space coordinates 
through the Hamiltonian flow, e.g., $A(t,\bm{z}) = \Phi_{t_0,t}^{*}A(t_0, \bm{z})$ where $\Phi_{t_0,t} = \Phi_{t,t_0}^{-1}$ is the backward ($t \to t_0$) map along the flow. 

The mathematical ``back-and-forth'' process provides a unique diagnostic that is inaccessible through standard methods. 
In contrast to conventional Eulerian correlation functions, which measure local fluctuations at a fixed point, 
or Lagrangian analysis, which tracks individual particle trajectories, 
the present classical-limit OTOC identifies the growth of the field response/sensitivity to initial perturbations 
across the entire phase space. 
The temporal reciprocity is fundamental to capture the essence of the quantum OTOC path, i.e., mapping from $t_0$ to $t$ and back to $t_0$, 
and is useful for characterizing the dynamical scrambling and the ``butterfly effect'' in classical turbulent fields.

\subsection{Extension to general non-canonical systems}
The derivation above utilized the advantage of canonical coordinates and the standard Moyal product on the microscopic phase space.
However, many classical field theories describing macroscopic dynamics of fluids and plasmas 
inherently possess non-canonical Poisson structures, where the dynamics is generated by a Lie-Poisson bracket
on an infinite-dimensional state space, i.e., Poisson manifold. 

To motivate an extension of Eq. (17) to more general non-canonical systems that subsume canonical ones, 
we recall a key mathematical result from the theories of deformation quantization and pseudo-differential operators.
Namely, for a general smooth Poisson manifold $(\mathcal{M}_{\mathcal{P}},\{\cdot,\cdot\}_{\mathcal{P}})$, 
there exists a star product deforming the commutative operator product such that its leading antisymmetric part reproduces 
the given Poisson bracket. 
This is guaranteed by Kontsevich's deformation-quantization theorem\cite{Kontsevich2003}. 

In formal expression, for a given Poisson bracket $\{\cdot,\cdot\}_{\mathcal{P}}$ on $\mathcal{M}_{\mathcal{P}}$ 
one can construct an associative star product $\circledast$: 
\begin{eqnarray}
\mathcal{F}\circledast \mathcal{G} = \mathcal{F}\mathcal{G} + \frac{i\hbar}{2}\left \{\mathcal{F},\mathcal{G} \right \}_{\mathcal{P}} + \mathcal{O}(\hbar^2), \nonumber \\
\mathcal{F}\circledast \mathcal{G} - \mathcal{G}\circledast \mathcal{F} = i\hbar\left \{\mathcal{F},\mathcal{G} \right \}_{\mathcal{P}} + \mathcal{O}(\hbar^2),
\end{eqnarray}
where $\mathcal{F}$ and $\mathcal{G}$ denote arbitrary functions (or functionals) on the Poisson manifold. 
Thus the corresponding ``Moyal-like'' bracket reduces to the Poisson bracket on $\mathcal{M}_{\mathcal{P}}$ at the lowest order of $\hbar$.
This argument provides the conceptual basis for defining a classical-limit OTOC for non-canonical Hamiltonian systems, 
i.e., we can define it by replacing the canonical Poisson bracket in Eq. (17) 
by the relevant non-canonical Poisson bracket $\{\cdot,\cdot\}_{\mathcal{P}}$.
Strictly speaking, the above existence result for a star product applies to finite-dimensional Poisson manifolds, 
so that the turbulence fields in mode-truncated representations fit exactly. 
For the infinite-dimensional Lie-Poisson structure with functionals, 
we do not claim the exact existence of a corresponding star product in full generality, 
but we can still construct a reasonable definition for the classical-limit OTOC as a turbulence diagnostic.  

A possible quantum counterpart of the present non-canonical OTOC can be formulated at the same algebraic level.
In an operator formulation, the classical observables or functionals are replaced by quantum operators.
Equivalently, in deformation quantization, these operators are represented by symbols, which are functions on the underlying Poisson manifold. 
In the present notation, such symbols are written as functionals, e.g., $\mathcal F[q]$, where $q$ denotes the field variable 
specifying a point on the Poisson manifold.
The ordinary product is then replaced by a noncommutative star product adapted to the non-canonical Lie-Poisson structure.
If such a star product is available, the corresponding non-canonical quantum OTOC can be formally represented by the star commutator as
\begin{equation}
C_{\mathcal F \mathcal G}^{\circledast_{\rm LP}}(t,t_0) 
= 
\left\langle 
[\mathcal F(t),\mathcal G(t_0)]_{\circledast}^{\dagger}
\circledast
[\mathcal F(t),\mathcal G(t_0)]_{\circledast}
\right\rangle ,
\nonumber
\end{equation}
where $ [\mathcal F,\mathcal G]_{\circledast} := \mathcal F\circledast\mathcal G - \mathcal G\circledast\mathcal F$. 
An adjoint operation for the star product is denoted by the dagger symbol, 
and $\langle\cdots\rangle$ stands for the expectation value with respect to the corresponding quantum state.
Since $(i\hbar)^{-1} [\mathcal F,\mathcal G]_{\circledast} = \{\mathcal F,\mathcal G\}_{P} + O(\hbar)$, 
the classical limit of $\hbar^{-2}C_{\mathcal F \mathcal G}^{\circledast_{\rm LP}}$ gives the non-canonical classical-limit OTOC used below.
The non-canonical quantum OTOC can be formulated naturally for finite-dimensional Poisson systems, e.g., mode-truncated fields.
A fully rigorous quantization of the infinite-dimensional Lie-Poisson field theory is beyond the scope of the present paper. 
Nevertheless, the above correspondence motivates the classical-limit OTOC proposed here as a useful diagnostic for H-M turbulence. 
%

\section{OTOCs for plasma and fluid turbulence}
In this section, we apply the classical-limit OTOC derived in the previous section to the Hasegawa-Mima (H-M) equation, 
which serves as a fundamental model for describing two-dimensional turbulence in magnetized plasmas. 
It is also of formal analogue to the 2-dimensional Euler system in neutral fluids. 
In general, fluid and plasma dynamics are difficult to formulate in terms of canonical variables. 
Therefore, we adopt the non-canonical Hamiltonian formalism on Poisson manifolds $\mathcal{M}_{\mathcal{P}}$ with a Lie-Poisson bracket, 
as established by Morrison et al. (see e.g., Refs. \cite{Morrison1982,Zyoshida2016}), to treat the field variables of the H-M system. 
A general theory for the H-M equation in non-canonical formulation\cite{Sato2024} is also related to the present approach. 
\subsection{Hasegawa-Mima equation for wave-vortex turbulence}
Let $\bm{x} = (x,y)$ be the 2-dimensional real-space coordinates, 
and we consider the dissipationless H-M equation for the electrostatic potential fluctuation $\phi(t,\bm{x})$ in the normalized form. 
The potential vorticity, which is also called the generalized vorticity, is defined as
\begin{equation}
q(t,\bm{x}) := \big(1-\nabla_{\perp}^2\big)\,\phi(t,\bm{x}) + \beta x,
\end{equation}
where $\nabla_{\perp}^2:=\partial_x^2+\partial_y^2$, and $\beta:= \partial_x \ln n_0$ denotes a parameter characterizing the background 
density inhomogeneity/gradient 
in the $x$-direction, i.e., $\beta < 0$ for a density profile decreasing in the positive x-direction. 
By using the potential vorticity, the governing equation is described as 
\begin{equation}
\partial_t q(t,\bm{x}) + \left \{\phi(t,\bm{x}), q(t,\bm{x})\right \}_{xy} = 0,
\end{equation}
where $\{f,g\}_{xy}:=\partial_x f\,\partial_y g-\partial_x g\,\partial_y f$ stands for the Poisson bracket in the 2-dimensional space $(x,y)$.  
Note also that the above H-M equation is formally analogous to the 2-dimensional Euler equation 
in the limit of $q = -\nabla_{\perp}^{2}\phi$.  

When we consider the potential vorticity as an infinite-dimensional state space, i.e., $q \in \mathcal{M}_{\mathcal{P}}$, 
the H-M equation in Eq. (20) is described as a non-canonical Hamiltonian system on the Poisson manifold. 
Suppose $\mathcal{F}[q(t,\cdot)] \in \mathbb{R}$ and $\mathcal{G}[q(t,\cdot)] \in \mathbb{R}$ 
be functionals of $q$, the Lie-Poisson bracket is defined by using the functional derivative $\delta/\delta q$ as
\begin{equation}
\left \{\mathcal{F},\mathcal{G} \right \}_{\rm HM}[q(t,\cdot)]
:=
\! - \! \int \! d\bm{x}\, q(t,\bm{x})\,
\left \{ \frac{\delta \mathcal{F}}{\delta q}[q(t,\cdot)]\, , \frac{\delta \mathcal{G}}{\delta q}[q(t,\cdot)] \right \}_{xy},
\end{equation}
where the time variable $t$ is treated as the fixed parameter. 
The Hamiltonian functional
\begin{eqnarray}
\mathcal{H}[q(t,\cdot)]
\!\! & := \!\! & 
\frac{1}{2}\! \int\! d\bm{x}\,\phi(t,\bm{x})\,\left (q(t,\bm{x})-\beta x\right) \nonumber \\
\!\! & = \!\! &
 \frac{1}{2}\! \int\! d\bm{x}\,\left(\left |\nabla_{\perp}\phi\right |^2+\phi^2\right),
\end{eqnarray}
where $\phi$ is obtained from $q$ via the inversion of $(1-\nabla_{\perp}^2)\phi=q-\beta x$.
Then the dynamical equation for the time evolution of an arbitrary functional $\mathcal{F}[q(t,\cdot)]$ is given by 
\begin{equation}
\partial_t \mathcal{F}[q(t,\cdot)] = \left \{\mathcal{F}[q(t,\cdot)],\mathcal{H}[q(t,\cdot)] \right \}_{\rm HM}.
\end{equation}
In particular, choosing $\mathcal{F}[q(t,\cdot)]= \! \int \! d\bm{x}\, \eta(\bm{x})q(t,\bm{x})$ with an arbitrary test function $\eta$ 
and using a weak-form derivation in Appendix A, Eq. (23) reproduces the H-M equation in Eq. (20). 
It is noted that Eq. (21) is a specific Lie-Poisson bracket for the H-M equation, 
involved in a special subclass of general Poisson brackets\cite{Morrison1982}. 

From the dynamical equation in Eq. (23), it follows immediately that the Hamiltonian is a conserved quantity. 
In addition to the energy, non-canonical Hamiltonian systems possess a set of conserved quantities $\mathscr{C}[q(t, \cdot)]$ 
known as Casimir invariants, which satisfy $\{\mathscr{C}, \mathcal{F}\}_{\rm HM} = 0$ for an arbitrary functional $\mathcal{F}[q]$. 
In the H-M system and 2D-Euler system, 
any functional of the form $\mathscr{C}[q(t, \cdot)] = \! \int d\bm{x} \, g(q(t,\bm{x}))$ with an arbitrary function $g$ 
serves as a Casimir invariant, implying that the fluid motion is geometrically constrained on a specific hypersurface called ``symplectic leaf'' 
where all Casimir functionals remain constant. 
Despite these infinite number of constraints, the H-M system still retains an infinite-dimensional degree of freedom, 
allowing for the emergence of complex turbulent flows and scrambling within the leaf in the manifold. 
%

\subsection{Classical-limit OTOC in H-M turbulence}
Let $A[q]$ and $B[q]$ be real-valued functionals (observables) of the state $q(t, \bm{x})$ in the H-M turbulence.
Motivated by Eq. (17) and the discussion of deformation quantization in Sec. 2.3, we define the classical-limit OTOC for the H-M turbulence as
\begin{equation}
C_{AB}^{\rm HM}(t,t_0)
:=
\left\langle \left (\{A[q(t,\cdot)],B[q(t_0, \cdot)]\}_{\rm HM}\right )^2\right\rangle_{\rm ens}. 
\end{equation}
Note that $t_0 (\le t)$ is arbitrary, e.g., either a linearly growing phase by instability or a nonlinearly saturated phase in developed turbulence. 
The ensemble average is taken over initial snapshots of turbulence fields at the reference time $t_0$, i.e., $q_0:=q(t_0)$.  
For statistically independent or sufficiently decorrelated snapshots $q_0^{(n)}=q^{(n)}(t_0)$, $n=1,\cdots,N$, we define 
$\langle X\rangle_{\rm ens}:=N^{-1}\sum_{n=1}^{N}X[q_0^{(n)}]$. 

A key point of Eq. (24) is the interpretation of the Lie-Poisson bracket $\{A[q(t,\cdot)],B[q(t_0,\cdot)]\}_{\rm HM}$ 
as an intrinsic variational response of a functional $A$ at time $t$, which is evolved by the Hamiltonian flow, 
to an initial perturbation along the flow induced by $B$ at time $t_0$.
In order to highlight such physical pictures, let us consider two flow maps coexisting independently. 
One is the time-evolution flow map generated by Eq. (23) with the Hamiltonian $\mathcal{H}$: 
\begin{equation}
q(t,\bm{x}) = \Phi_{t,t_0}q(t_0, \bm{x}), 
\end{equation}
where $\Phi_{t,t_0}$ means the forward flow map. Note that $\Phi_{t_0,t_0} = \rm id$. 

Independently of the physical time $t$, 
we introduce a one-parameter ($\eps$) family of perturbations generated by an arbitrary functional $B[q]$.
Specifically, for any smooth functional $F[q]$, a flow map $T_\eps$ is defined as a solution of 
\begin{equation}
d_\veps F\!\left[T_\eps q\right]
=
\{F,B\}_{\rm HM}\!\left[T_\eps q\right],
\end{equation}
where $d_{\veps}:= d/d\veps$ and $T_0 = \rm id$. Note also that $T_{\eps}$ determines another flow map generated by $B$, not by $\mathcal{H}$. 
Then one can define a perturbed initial state as $q^{(\epsilon)}(t_0,\bm{x}) = T_{\eps}q(t_0,\bm{x})$. 
Combining the above two flow maps, the time-evolved perturbed state is given by 
\begin{equation}
q^{(\eps)}(t,\bm{x})
:=
\Phi_{t,t_0}q^{(\eps)}(t_0,\bm{x}) = \Phi_{t,t_0} \circ T_{\eps}\left [q(t_0, \bm{x}) \right ]. 
\end{equation}

Once the time-evolved state perturbed along the vector field generated by $B[q]$ at time $t_0$ is determined, 
the variational response of an arbitrary functional of $A[q]$ at later time $t$ is calculated by G\^{a}teaux derivative: 
\begin{equation}
\delta A(t\,|\,t_0, B) := \left.d_\veps A\!\left[q^{(\veps)}(t, \cdot) \right ] \right|_{\veps=0} 
= \left. d_{\veps} T_{\veps}^{\ast}\Phi_{t,t_0}^{\ast} A\!\left[q(t_0, \cdot)\right] \right|_{\veps=0}, 
\end{equation}
where the superscript of $\ast$ on $\Phi_{t,t_0}$ and $T_{\eps}$ means the pull-back map acting on the functional $A$, 
i.e., $(\Phi_{t,t_0} \! \circ T_{\eps})^{\ast}A = T_{\eps}^{\ast}\Phi_{t,t_0}^{\ast}A$.  
After some algebra using Eq. (26), i.e., applying $F[q] = \Phi_{t,t_0}^{\ast}A[q(t_0, \cdot)]$ and evaluate it at $q=q(t_0, \cdot)$, 
one can derive the following relation: 
\begin{equation}
\left. d_{\veps}
T_{\veps}^{\ast}\Phi_{t,t_0}^{\ast}A\!\left[q(t_0, \cdot)\right]
\right|_{\veps=0}
=
\left.
\left\{
\Phi_{t,t_0}^{\ast}A\!\left[q\right],
B[q]
\right\}_{\rm HM}
\right|_{q=q(t_0, \bm{x})}.
\end{equation}
It should be emphasized that since the specific form of $\{\cdot ,\cdot \}_{\rm HM}$ is not used in the derivation, 
Eq. (29) holds for the general canonical and non-canonical systems in Poisson manifolds.     

Now we can combine the initial perturbation generated by $B$, the variational response $\delta A$, and the classical-limit OTOC 
$\mathcal{C}_{AB}^{\rm HM}(t,t_0)$ in the non-canonical H-M system, as follows: 
\begin{equation}
\left\langle \left (\delta A(t\,|\,t_0, B) \right )^2 \right\rangle_{\rm ens} 
= \left\langle \left ( \left. d_{\veps} T_{\veps}^{\ast}\Phi_{t,t_0}^{\ast}A\!\left[q(t_0, \cdot)\right] \right|_{\veps=0}  \right )^2 \right\rangle_{\rm ens}
=  \mathcal{C}_{AB}^{\rm HM}(t,t_0) 
\end{equation}
This identity clarifies the physical picture and usefulness of the classical-limit OTOC in the H-M turbulence and the other related fluid systems. 
While standard linear response theory, e.g., Green-Kubo formula\cite{GKubo}, characterizes macroscopic transport via 
the mean response function $L_{AB}(t) := \left\langle \{ A(t), B(t_0) \}_{\rm PB} \right\rangle_{\rm ens}$ for the externally 
forced perturbation in canonical systems, 
it often fails to capture chaotic information due to the phase cancellation among microscopic trajectories. 
In turbulent flows, the sign of the Lie-Poisson bracket fluctuates rapidly across the phase space, 
so that the mean response $L_{AB}(t)$ vanishes or decays in time.
Without such cancellation, the classical-limit OTOC $\mathcal{C}_{AB}^{\rm HM}(t,t_0)$ 
can measure the scrambling process (or butterfly effect) in terms of the growth of quadratic nonlocal response between 
arbitrary physical quantities (functionals) $A$ and $B$. 

Figure 1 shows the schematic diagram of the classical-limit OTOC as a quadratic variational response to initial perturbations. 
One can see, from the figure, geometric concepts of the OTOC with two different flow maps along the vector fields generated by $\mathcal{H}[q]$ and $B[q]$. 
\begin{figure}[h]
\centering
\includegraphics[scale=0.25]{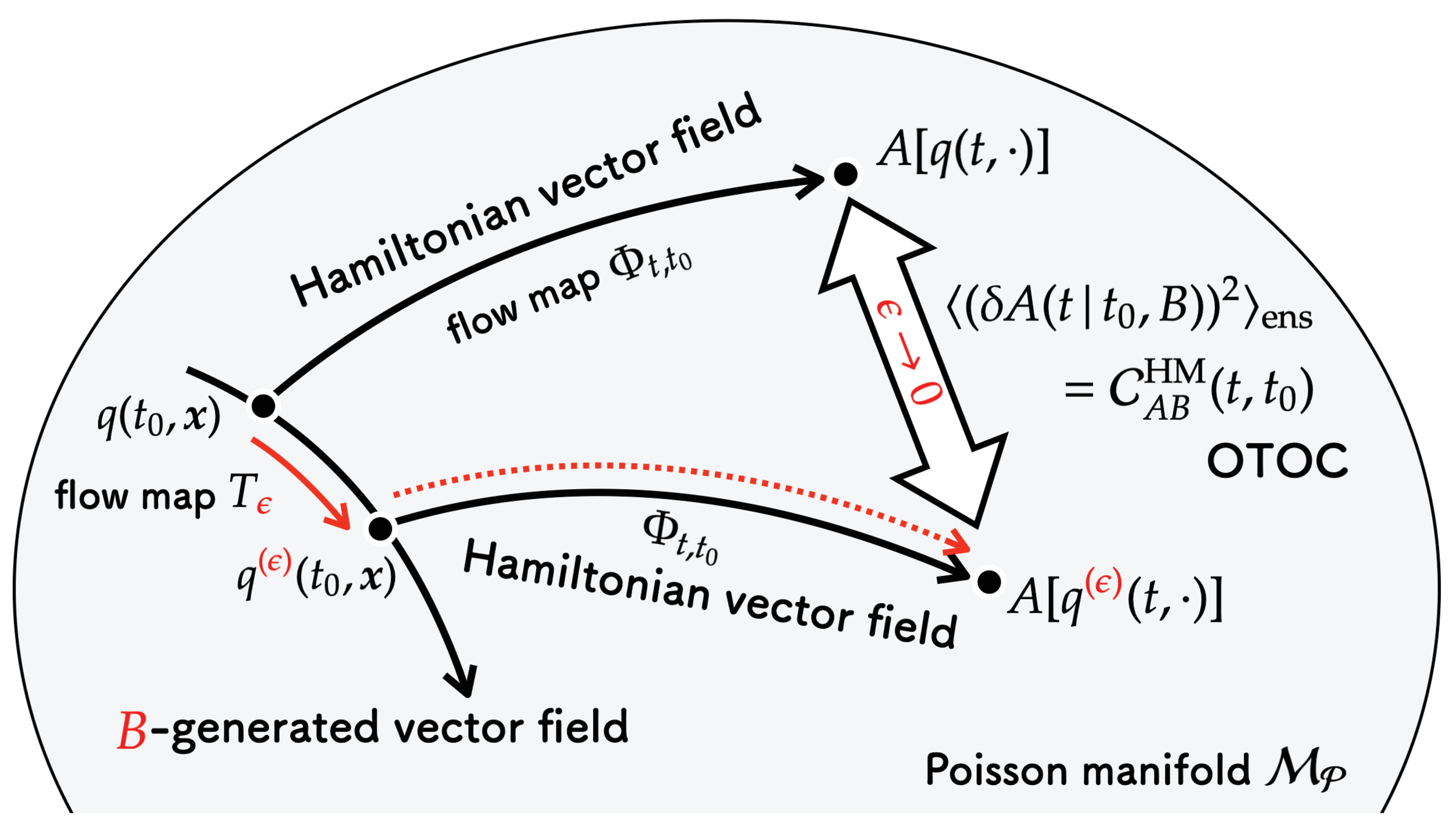}
\caption{Schematic diagram of the classical-limit OTOC as a quadratic variational response to initial perturbations.}
\end{figure}

As for the specific form of perturbations represented by $T_{\veps}$, one can easily find 
\begin{equation}
\delta q_B(t_0,\bm{x})
:=
\left. d_{\veps}q^{(\veps)}(t_0,\bm{x}) \right |_{\veps=0} 
=
\{q(t_0, \bm{x}),B\}_{\rm HM}
= \left. \left \{q, \frac{\delta B}{\delta q} \right \}_{xy} \right |_{q=q(t_0,\bm{x})}, 
\end{equation}
where the derivation is shown in Appendix B. 
Note that $\delta q_B$ means a perturbed state along a vector field on the Poisson manifold generated by the functional $B[q]$. 

After establishing Eq. (30), we can introduce complementary projectors $P$ and $Q:=1-P$
to define the relevant physical quantities, $q_P:=Pq$ and $q_Q:=Qq$, for subsystems interacting each other, 
e.g., multiple components, domains, and scales. 
Then $A_P=A[q_P]$ and $B_Q=B[q_Q]$ are chosen to quantify a specific spatio-temporal response to the perturbation, 
e.g., non-zonal to zonal fields shown in the next section. 
%

\subsection{Scaling of OTOC for zonal-mode response to non-zonal perturbation}
We now clarify the physical meaning of the classical-limit OTOC for two turbulent fields (zonal and non-zonal modes) in the H-M turbulence.  
The goal is to obtain an analytic expression of the classical-limit OTOC in Eq. (30) as the quadratic intrinsic response 
of a large-scale zonal mode at time $t$ to an initially imposed fine-scale non-zonal perturbation at time $t_0$. 
Then, the asymptotic scalings with respect to the time lag $\Delta t:=t-t_0$ in short-time and long-time limits are evaluated 
under several approximations. 
Note that we consider $t_0$ as a specific time in the fully-developed turbulence state, not the initial quiescent state. 

Let $P$ be the zonal projection as the $y$-average,
\begin{equation}
Pf(t,x):=\left \langle f(t,\bm{x}) \right \rangle_y = \frac{1}{L_y} \! \int \! dy f(t,\bm{x}),
\end{equation}
and $Q:=1-P$ denotes the non-zonal projection. The system size in the $y$-direction is described by $L_y$.   
The fields are then decomposed into the sum of zonal modes and non-zonal modes, i.e.,  
$f=\bar{f}+\tilde{f}$, $\bar{f}:=Pf$, $\tilde{f}:=Qf$. 
Since the background gradient term $\beta x$ appears in $\bar{q}$, 
the non-zonal potential vorticity is simply written as $\tilde{q}=(1-\nabla_{\perp}^2)\,\tilde{\phi}$ because of $Q[\beta x]=0$. 

We choose $B[q]$ as a functional that injects a non-zonal perturbation localized in a targeted wavenumber range around 
$\bm{k}_{0} = (k_{x0}, k_{y0})$ through a kernel $\chi(\bm{x};\bm{k}_0)$: 
\begin{equation}
B[q]:= Q_{\rm NZ}(t_0;\bm{k}_0) = \! \int\! \dd \bm{x}\,\chi(\bm{x};\bm{k}_0)\,\tilde q(t_0, \bm{x}),
\end{equation}
where the condition $P\chi = \langle\chi\rangle_y=0$ is imposed. 
Although this condition is not strictly required because $\tilde q$ is already non-zonal by definition, it simplifies the functional derivative expression. 
A convenient example of $\chi$ is a Gaussian kernel in Fourier form:
\begin{equation}
\chi(\bm{x};\bm{k}_{0})= \frac{1}{(2\pi)^2} \! \int \! d\bm{k}\,
\exp\!\left(-\frac{(k_x-k_{x0})^2}{2\Delta k_x^2}-\frac{(k_y-k_{y0})^2}{2\Delta k_y^2}\right)
\exp (i\bm{k}\cdot\bm{x}).
\end{equation}

By using Eq. (31) with $\delta B/\delta q=\chi-P\chi = \chi$, 
The induced initial perturbation generated by $B[q]$ is written as 
\begin{equation}
\delta q_B(t_0,\bm{x})
=
\{q(t_0,\bm{x}),B\}_{\rm HM}
=
\left \{\,q(t_0,\bm{x}),\chi(\bm{x};\bm{k}_0)\,\right \}_{xy}, 
\end{equation}
where the non-zonal projection is given by $\delta\tilde q_B(t_0,\bm{x})=Q\,\delta q_B(t_0,\bm{x})$. 

As for the variational response of $A$, we consider the energy of large-scale zonal modes:   
\begin{equation}
A[q]:=E_{\rm ZF}(t)=\frac{1}{2}\! \int \! d\bm{x}\,U(t,x)^2, 
\end{equation}
where $U(t,x):=-\partial_x \bar{\phi}(x,t)$ means the zonal flow velocity. 
Then the classical-limit OTOC for non-zonal to zonal response can be written as 
\begin{equation}
C_{AB}^{\rm HM}(t,t_0)
=
\left\langle \delta E_{\rm ZF}(t\,|\,t_0,Q_{\rm NZ})^2\right\rangle_{\rm ens}. 
\end{equation}
Note that, in addition to the initial snapshots of $q(t_0)$, variations of the kernel $\chi$ can also be included in the ensemble average.
For instance, if $\chi^{(m)}$, $m=1,\cdots,M$, denote different choices of the kernel phase, center, width, or target wavenumber, 
the ensemble average can be extended as $\langle X\rangle_{\rm ens}:=(NM)^{-1}\sum_{n=1}^{N}\sum_{m=1}^{M}X[q_0^{(n)},\chi^{(m)}]$.
The variational response of $E_{\rm ZF}$ to the initial perturbation induced by $Q_{\rm NZ}$ is calculated by 
\begin{equation}
\delta E_{\rm ZF}(t\,|\,t_0,Q_{\rm NZ})
:=\left.d_\veps E_{\rm ZF}[q^{(\veps)}(t)]\right|_{\veps=0}
=
\! \int \! d\bm{x}\ U(t,x)\,\delta U(t,x),
\end{equation}
where $\delta U:=\left.d_\veps U^{(\veps)}\right|_{\veps=0}$ 
denotes the G\^{a}teaux derivative along $B[q] = Q_{\rm NZ}$. 
Since $\delta E_{\rm ZF}$ is linear in $\delta U$, estimating the time-lag scaling of $C_{AB}^{\rm HM}$ 
reduces to estimating the scaling of $\delta U$ based on the H-M equation, 
and different scaling in the short-time or long-time limit is shown in the following. 

Applying the projectors $P$ and $Q$ to the H-M equation in Eq. (20) with the perturbations of $\delta q_B$ and $\delta \phi_B$, 
one obtains the dynamical equations for the non-zonal modes:  
\begin{eqnarray}
\partial_t \delta \tilde{q}_B
 \! \! &+& \! \! \big\{\bar{\phi},\delta \tilde{q}_B\big\}_{xy} +\big\{\tilde{\phi},\delta \bar{q}_B\big\}_{xy} +Q\big\{\tilde{\phi},\delta \tilde{q}_B\big\}_{xy} \nonumber \\
 \! \! &+& \! \! \big\{\delta \bar{\phi}_B,\tilde{q}\big\}_{xy} +\big\{\delta \tilde{\phi}_B,\bar{q}\big\}_{xy} +Q\big\{\delta \tilde{\phi}_B,\tilde{q}\big\}_{xy} \nonumber \\
 \! \! &+& \! \! \big\{\delta \bar{\phi}_B,\delta \tilde{q}_B\big\}_{xy} +\big\{\delta \tilde{\phi}_B,\delta \bar{q}_B\big\}_{xy} +Q\big\{\delta \tilde{\phi},\delta \tilde{q}_B\big\}_{xy} = 0.  
\end{eqnarray}
Here we impose the quasilinear approximation and the first variation ansatz, 
i.e., the products of non-zonal modes and of $\delta q_B$ and $\delta \phi_B$ are ignored as more-than-the-second-order effects. 
The non-zonal mode equation is, thus, reduced to the leading-order tangent linear equation with respect to the perturbation:       
\begin{equation}
\partial_t \delta \tilde{q}_B - U(t,x)\partial_y \delta \tilde{q}_B - G(t,x) \partial_y \delta \tilde{\phi}_B = 0, 
\end{equation}
where $\{\bar{\phi},\delta \tilde{q}_B\}_{xy} = - U(t,x)\partial_y \delta \tilde{q}_B$ 
and $\{\delta \tilde{\phi}_B,\bar{q}\}_{xy} = - G(t,x) \partial_y \delta \tilde{\phi}_B$ with $G(t,x):= \partial_x \bar{q}$ are used. 
For more general argument of the linear tangent equation, see Appendix C. 
Note also that $\delta \tilde{q}_B = (1-\nabla_{\perp}^2)\delta \tilde{\phi}_B$ from Eq. (19). 

Since Eq. (40) is a linear advection equation, 
the local co-moving coordinates $x^{\prime} := x-x_0$, $y^{\prime}:= y + \int_{t_0}^{t} d\tau \{ U_0(\tau) + S(\tau)(x-x_0)\}$ yields 
\begin{equation}
\partial_t \delta \tilde{q}_B(t,x^{\prime}, y^{\prime}) = G_0 \partial_{y^{\prime}} \delta \tilde{\phi}_B(t,x^{\prime}, y^{\prime}), 
\end{equation}
where $U(t,x) \simeq U_0(t) + S(t)(x-x_0)$ with $U_0:= U(t,x_0)$ and $G_0 := G(t,x_0)$.  
The Fourier expansion for $\delta \tilde{q}_B$ gives 
\begin{eqnarray}
\delta\tilde{q}_B
 \! \! & = &  \! \! 
\sum_{\bm{k}}^{\rm (nz)} \widehat{\delta \tilde{q}}_{B\bm{k}}\,
\exp\!\big[i(k_{x0}x^{\prime}+k_y y^{\prime})\big]  \nonumber \\ 
 \! \! &=&  \! \! \sum_{\bm{k}}^{\rm (nz)} \widehat{\delta \tilde{q}}_{B\bm{k}}\exp\!\big[i(k_x(t)x^{\prime}+k_y y)\big]\,
\exp\!\left[ik_y \! \int_{t_0}^{t} \! d\tau\, U_0(\tau)\right],  \\
k_x(t)  \! \! & := &  \! \! k_{x_0} + k_y \int_{t_0}^{t} \! d\tau\, S(\tau), 
\end{eqnarray}
where $\sum_{\bm{k}}^{\rm (nz)}$ denotes the summation over the non-zonal wavenumber modes with $\bm{k}=(k_{x0},k_y)$.  
Note that the hat symbol in this subsection means the Fourier coefficient, not the operator. 
Using the time-dependent wavenumber $k_x(t)$, one obtains 
\begin{eqnarray}
\widehat{\delta \tilde{\phi}}_{B\bm{k}}(t) = (1+k_x^2(t) + k_y^2)^{-1}\, \widehat{\delta \tilde{q}}_{B\bm{k}}(t), \\
d_t \widehat{\delta \tilde{q}}_{B\bm{k}}(t) - ik_y G_0(1+k_x^2(t) + k_y^2)^{-1}\,\widehat{\delta \tilde{q}}_{B\bm{k}}(t) = 0, 
\end{eqnarray}
leading straightforwardly to $|\delta\tilde{q}_B(t)|^2 = |\delta\tilde{q}_B(t_0)|^2$. 
A similar argument holds for $|\tilde{q}(t)|^2 = |\tilde{q}(t_0)|^2$, but 
note that $|\delta\tilde{\phi}_B(t)|^2 \neq |\delta\tilde{\phi}_B(t_0)|^2$ and $|\tilde{\phi}(t)|^2 \neq |\tilde{\phi}(t_0)|^2$ 
because $|k_x(t)|$ grows asymptotically in time due to the shearing effect. 
This continuous deformation of the wavevector leads to an algebraic decay of the amplitude of $\delta\tilde{\phi}_B$ and $\tilde{\phi}$, 
which is also known as the rapid distortion theory (RDT) in fluid dynamics.  

As for the zonal modes, the time evolution is determined by the higher-order correlations:    
\begin{equation}
\partial_t \delta \bar{q}_B 
+ P\left [ \big\{\tilde{\phi},\delta \tilde{q}_B\big\}_{xy} + \big\{\delta \tilde{\phi}_B,\tilde{q}\big\}_{xy} 
+ \big\{\delta \tilde{\phi}_B,\delta \tilde{q}_B\big\}_{xy} \right ] =0. 
\end{equation}
Focusing on the first-order variational response of $\delta \bar{q}_B$ by neglecting $\{\delta \tilde{\phi}_B,\delta \tilde{q}_B\}_{xy}$, we then integrate Eq. (46) as follows: 
\begin{eqnarray}
\delta \bar{q}_B(t) = - \! \int_{t_0}^{t} \! d\tau\,   
P\left [ \big\{\tilde{\phi},\delta \tilde{q}_B\big\}_{xy} + \big\{\delta \tilde{\phi}_B,\tilde{q}\big\}_{xy} \right ], 
\end{eqnarray}
where $\delta \bar{q}_B(t_0) = 0$ by definition.
Note that the contribution of $\{\delta \tilde{\phi}_B,\delta \tilde{q}_B\}_{xy}$ is one-order smaller, 
so that its neglect does not matter to evaluating $\Delta t$ scaling. 
By using Eqs. (44)--(45) and (47) with the Fourier representation in Eqs. (42) and (43), 
we can derive an analytic expression of the zonal-flow-energy response $\delta E_{\rm ZF}(t\,|\,t_0,Q_{\rm NZ})$ to calculate the OTOC in Eq. (37): 
\begin{equation}
\delta E_{\rm ZF}(t\,|\,t_0,Q_{\rm NZ})
=
L_xL_y \sum_{k_{\rm zf}} \frac{k_{\rm zf}^2}{(1+k_{\rm zf}^2)G_0} \widehat U_{-k_{\rm zf}}(t) \sum_{\bm{k}}^{({\rm nz})}
\mathcal P_{\bm{k}}(t_0)
\left\{
\exp\!\left[i k_yG_0\mathcal I_{\bm{k}}(t,t_0)\right]-1
\right\}, 
\end{equation}
where 
\begin{eqnarray}
\mathcal P_{\bm{k}}(t) 
 \! &=& \!
\widehat{\tilde q}_{\bm{k}}(t)\, \widehat{\delta\tilde q}_{B\bm{k}_{\rm zf}-\bm{k}}(t), \\
\mathcal I_{\bm{k}}(t,t_0)
 \! &=& \!
\frac{1}{k_yS_0K} \Bigg[ \tan^{-1}\!\left( \frac{k_{x0}+k_yS_0(t-t_0)}{K} \right) - \tan^{-1}\!\left( \frac{k_{x0}}{K} \right) \nonumber \\
&&
- \tan^{-1}\!\left( \frac{k_{\rm zf}-k_{x0}}{K} \right) + \tan^{-1}\!\left( \frac{k_{\rm zf}-k_{x0}-k_yS_0(t-t_0)}{K} \right) \Bigg]. 
\end{eqnarray}
Here we also used the expression of $U(t,x^{\prime})= \sum_{k_{\rm zf}} \widehat U_{k_{\rm zf}}(t)\exp[i k_{\rm zf}x^{\prime}]$ 
with the wavenumber for the zonal modes $\bm{k}_{\rm zf} = (k_{\rm zf},0)$, and $K:=(1+k_{y}^2)^{\frac{1}{2}}$. 
More detailed derivations are given in Appendix D.  
It is emphasized that Eqs. (37) and (48)--(50) provide a closed analytic expression of the classical-limit OTOC $C_{AB}^{\rm HM}(t,t_0)$ 
for the H-M turbulence within the local quasilinear and constant flow-shear ansatz.
In deriving the asymptotic estimates below, we additionally assume that the background zonal-flow amplitude $\widehat U_{k_{\rm zf}}(t)$ varies slowly 
over the time scale of non-zonal modes, and that the strong-shear ordering for $S_0$ is applicable for the long-time limit estimate.

Based on the above analytic expression, two asymptotics for the OTOC are evaluated, 
i.e., short-time limit for $\Delta t \ll 1$ and long-time limit for $\Delta t \gg 1$. 
The detailed calculations are left to the latter part in Appendix D, and the results are summarised here: 
\begin{equation}
  C_{AB}^{\rm HM}(t,t_0) \simeq
  \begin{cases}
    \ \left\langle \mathcal A(t_0)^2\right\rangle_{\rm ens} (\Delta t)^2 +O((\Delta t)^3) & (\Delta t \ll 1), \\
    \ C_{AB}^{\rm HM}(\infty,t_0) + O\left[S_0^{-4}(\Delta t)^{-2}\right] & (\Delta t \gg 1),
  \end{cases}
\end{equation}
where $\mathcal{A}(t_0)$ and $C_{AB}^{\rm HM}(\infty,t_0)$ are coefficients that do not depend on $\Delta t$, and their expressions are given in Appendix D.   
The short-time branch in Eq. (51) follows from a Taylor expansion in $\Delta t$.
For the long-time branch, we assume $\widehat U_{-k_{\rm zf}}(t) \simeq \widehat U_{-k_{\rm zf}}(t_0)$ and the strong-shear ordering $|G_0/S_0| \ll 1$.
It is found that the OTOC grows quadratically with respect to $\Delta t$ in the short-time limit, 
whereas it asymptotically approaches a finite saturated value with an inverse-square algebraic dependence in the long-time limit. 
The leading short-time growth coefficient $\left\langle \mathcal A(t_0)^2\right\rangle_{\rm ens}$ has no explicit $S_0$ dependence. 
This reflects the fact that, at sufficiently short time lags, the response is governed by the non-zonal-to-zonal coupling at $t_0$, 
before the shearing-induced growth of the wavenumber $k_x(t)$ becomes effective.
The crossover time from the short-time growth to the saturated regime can be defined by matching the short-time behavior to the saturated value, as follows: 
\begin{equation}
\tau_{\rm sat} := \left[ \frac{C_{AB}^{\rm HM}(\infty,t_0)} {\left\langle \mathcal A(t_0)^2\right\rangle_{\rm ens}} \right]^{1/2} = O(|S_0|^{-1}).
\end{equation}
These analytic results also provide a useful reference for numerical studies with the OTOC including fully nonlinear turbulent effects.

The algebraic dependence in Eq. (51) should be understood as a consequence of scale-to-scale transfer/redistribution of the perturbation 
rather than a direct decay of the amplitude, in line with the physical picture discussed in earlier work\cite{Nakata1, Nakata2}.
A perturbation injected into the non-zonal degrees of freedom by $\chi(\bm{x};\bm{k}_0)$ is rapidly sheared and mixed by the strong zonal flow, 
thereby transferring its spectral fluctuations toward higher wavenumber region. 
This process reduces the low-wavenumber non-zonal content that can feed back onto the large-scale zonal modes, 
and the resulting weakening of the residual zonal-mode response is captured by the long-time scaling in Eq. (51).
Thus, by constructing relevant projection operators for the functionals, the OTOC quantifies the time-dependent transfer 
of a perturbation between selected scales or field components of the turbulent fields. 

It is also noted that the absence of an exponential growth in Eq. (51) does not imply that a classical-limit OTOC in a non-canonical system 
cannot exhibit Lyapunov-type growth. 
If the tangent dynamics on a symplectic leaf specified by the Casimir invariants, or in a finite-mode truncation, has an unstable direction, 
a relevant choice of functionals $A$ and $B$ can project the non-canonical Lie-Poisson OTOC onto that direction. 
In such cases a Lyapunov-like exponential growth may appear, in close analogy with the canonical semiclassical OTOC theory\cite{Jalabert2018,Michel2025}.
On the other hand, in the present H-M turbulence, $A=E_{\rm ZF}$ and $B=Q_{\rm NZ}$ are chosen to diagnose the non-zonal-to-zonal functional response 
in the strong-shear quasilinear regime. 
This observable pair does not probe the maximal tangent-space instability directly, but instead it reveals the growth-to-saturation behavior summarized in Eq. (51).

\section{Concluding remarks}
In this paper we have developed a route to introduce quantum out-of-time-ordered correlators (OTOCs) into classical turbulence dynamics 
by making the quantum-classical correspondence explicit. 
In the semiclassical limit calculations based on Wigner-Weyl transform and Moyal bracket, 
the OTOC reduces to the ensemble average of a squared canonical Poisson bracket.
We have also shown a natural extension of the classical-limit OTOC for non-canonical Hamiltonian systems with a Lie-Poisson bracket, 
such as fluid and plasma turbulence, 
motivated by the deformation-quantization theorem for a general Poisson structure. 
The classical-limit OTOC can be understood as a useful measure of how a variational perturbation injected along a chosen functional $B$ at time $t_0$ 
propagates and influences another functional $A$ at later time. 
This allows us to construct scale-selective or field-selective diagnostics through projection operators, and is not restricted to a global Lyapunov exponent. 

As a concrete example we formulated a classical-limit OTOC for wave-vortex turbulence governed by the Hasegawa-Mima equation. 
Here the zonal/non-zonal decomposition in the quasilinear approximation is considered to evaluate the non-zonal to zonal mode response 
in a regime of strong zonal-flow shearing.
It has been clarified from an analytic expression that the OTOC grows quadratically, $O[(\Delta t)^2]$, at early time, 
whereas in the long-time strong-shear regime it approaches a finite saturated value with an inverse-square algebraic dependence, \(O[S_0^{-4}(\Delta t)^{-2}]\).
The saturation behavior should be interpreted in terms of shear-induced scale-to-scale transfer of non-zonal perturbations, rather than as a decay of the OTOC itself. 
Indeed, the non-zonal perturbation is rapidly scrambled toward higher wavenumbers by the
zonal-flow shear, and consequently the low-wavenumber non-zonal content capable of coupling back to the large-scale zonal modes is attenuated at later time. 

Several extensions are left for future work. 
First, beyond the quasilinear analytical scaling presented here, more quantitative numerical evaluations of the OTOC to include non-zonal eddy-eddy interactions 
will clarify how the inverse-square approach to saturation is modified by fully nonlinear turbulent mixing.

Second, it would be interesting to apply the OTOC analysis to multi-field fluid systems with finite dissipation 
such as Hasegawa-Wakatani turbulence (e.g., Ref. \cite{Guillon2025}), or phase-space kinetic systems such as gyrokinetic turbulence (e.g., Ref. \cite{Nakata3}).  

Third, another related direction is to examine whether a non-canonical Lie-Poisson OTOC can exhibit exponential growth 
near locally unstable equilibria or coherent structures on a symplectic leaf of the system. 
Indeed, in canonical quantum systems, exponential growth of the OTOC has been reported even in integrable or non-chaotic systems when the initial state is localized 
near an unstable fixed point or a locally hyperbolic structure\cite{Pappalardi2018,Hummel2019,Rozenbaum2020,Hashimoto2020,Steinhuber2023,Michel2026}.

The non-canonical classical-limit OTOC analysis has been demonstrated for the H-M turbulence, which is one of the most fundamental yet universal models. 
Introducing a perspective of the quantum-classical correspondence through the Weyl-Wigner-Moyal formalism has broader implications beyond a specific turbulence diagnostic.
A quantum counterpart of the present OTOC would require a quantization of the corresponding non-canonical Lie-Poisson algebra through a star product 
in a finite-mode truncation, and may provides useful theoretical insights toward the future realization of quantum algorithms and quantum computing for plasma turbulence. 

\newpage
\appendix
\section*{Appendices}
\setcounter{equation}{0}
\renewcommand{\theequation}{A\arabic{equation}}

\section{A. Non-canonical formulation of the H-M equation}
For clarity, we summarize the weak-form derivation that connects Eqs. (21) and (22) 
to the local PDE in Eq. (20) for the H-M turbulence.

Let $\eta(\bm{x})$ be an arbitrary smooth test function, and consider the linear functional 
\begin{equation}
\mathcal{F}_\eta[q(t,\bm{x})]:= \! \int \! d\bm{x}\, \eta(\bm{x})\,q(t,\bm{x}).
\end{equation}
By definition of the functional derivative,
\begin{equation}
\delta \mathcal{F}_\eta = \! \int \! d\bm{x}\, \eta(\bm{x})\,\delta q(t,\bm{x})
=
\! \int \! d\bm{x} \frac{\delta F_\eta}{\delta q(t, \bm{x})}\,\delta q(t,\bm{x}),
\end{equation}
where $\delta \mathcal{F}_\eta/\delta q=\eta(\bm{x})$.
Similarly, from Eq. (22) one obtains
\begin{equation}
\frac{\delta \mathcal{H}}{\delta q(t,\bm{x})}
= \frac{1}{2} \frac{\delta}{\delta q} \! \int \! d\bm{x}\, \phi\, (q-\beta x)
=\phi(t,\bm{x}),
\end{equation}
where $(1-\nabla_{\perp}^2)\,\phi=q-\beta x$ by the definition of $q(t,\bm{x})$. 
By using the dynamical equation in Eq. (23) with $\mathcal{F}=\mathcal{F}_\eta$, 
\begin{equation}
\partial_t \mathcal{F}_\eta[q(t,\cdot)]
=\{\mathcal{F}_\eta,\mathcal{H}\}_{\rm HM}[q(t,\cdot)]
= \! - \! \int \! d\bm{x}\, q(t,\bm{x})\left \{\eta(\bm{x}),\phi(t,\bm{x}) \right \}_{xy}.
\end{equation}
Integrating by parts and using antisymmetry of the Poisson bracket $\{\cdot,\cdot\}_{xy}$ under $\int \! d\bm{x}$ 
assuming periodic or vanishing boundaries, we rewrite Eq. (A4) as follows:
\begin{equation}
\int \! d\bm{x}\, \eta(\bm{x})\,\partial_t q(t,\bm{x})
=
- \! \int \! d\bm{x}\, \eta(\bm{x})\left \{\phi(t,\bm{x}),q(t,\bm{x}) \right \}_{xy}.
\end{equation}
Since $\eta(\bm{x})$ is arbitrary, we find the local PDE for the H-M turbulence of Eq. (20) in the weak-form sense: 
\begin{equation}
\partial_t q(t,\bm{x}) + \left \{\phi(t,\bm{x}), q(t,\bm{x}) \right \}_{xy} = 0.
\end{equation}
The same result is also derived by replacing 
$\int \! d\bm{x}\, \eta(\bm{x})\, q(t,\bm{x}) \to  \int \! d\bm{x} \,\delta(\bm{x}-\bm{x}_{0})\,q(t,\bm{x})$ in Eq. (A1) for an arbitrary 
fixed point $\bm{x}_0$. 
%

\section{B. Weak-form derivation of the functional-generated perturbation}
\setcounter{equation}{0}
\renewcommand{\theequation}{B\arabic{equation}}
In Sec.3.2 we used the short identification $\delta q_B(t_0,\bm{x})=\{q(t_0, \bm{x}),B\}_{\rm HM}$ in Eq. (31).  
Since $q(t,\bm{x})$ is a field (not a regular functional), we justify this identity in a weak form.

In the same manner of Eq. (A1), substituting $\mathcal{F}_\eta$ to Eq. (26) yields  
\begin{equation}
d_{\veps}\mathcal{F}_\eta[T_\veps q]
= \{\mathcal{F}_\eta,B\}_{\rm HM}[T_\veps q].
\end{equation}
Evaluating at $\veps=0$ gives
\begin{equation}
\left.d_{\veps}\mathcal{F}_\eta[T_\veps q]\right|_{\veps=0}
=
\{\mathcal{F}_\eta,B\}_{\rm HM}[q].
\end{equation}
Then, the left-hand side is calculated as 
\begin{equation}
\left.d_{\veps}\mathcal{F}_\eta[T_\veps q]\right|_{\veps=0}
=
\! \int \! d\bm{x}\, \eta(\bm{x}) \left.d_{\veps}T_\veps q(t,\bm{x})\right|_{\veps=0}
=
\! \int \! d\bm{x}\,\eta(\bm{x})\,\delta q_B(\bm{x}),
\end{equation}
which defines $\delta q_B$ in the weak sense.
As for the Lie-Poisson bracket on the right-hand side, we obtain 
\begin{equation}
\{\mathcal{F}_\eta,B\}_{\rm HM}[q]
=
 - \! \int \! d\bm{x}\, q(t,\bm{x})\,
\{\eta(\bm{x}),\gamma(t,\bm{x}) \}_{xy}, 
\end{equation}
where $\gamma(t,\bm{x}):= \delta B/\delta q(t,\bm{x})$. 

Similarly to Eq. (A5), $q(t,\bm{x})$ and $\eta(\bm{x})$ are exchanged 
by integration by parts in periodic boundaries or boundary conditions that vanish surface terms: 
\begin{equation}
\! \int \! d\bm{x}\, q(t,\bm{x})\,
\{\eta(\bm{x}),\gamma(t,\bm{x}) \}_{xy}
= 
- \! \int \! d\bm{x}\,\eta(\bm{x})\{q(t,\bm{x}),\gamma(t,\bm{x}) \}_{xy}
\end{equation}
Since $\eta$ is arbitrary, we identify the strong form relation by combining Eq. (B3) and (B5) as follows: 
\begin{equation}
\delta q_B(t,\bm{x}) = \{q(t, \bm{x}),B\}_{\rm HM} 
= \left \{q(t,\bm{x}), \frac{\delta B}{\delta q(t,\bm{x})} \right \}_{xy}. 
\end{equation}
For Eq. (31) in Sec. 3.2, this is evaluated at the initial state $q=q(t_0,\bm{x})$.

\section{C. Tangent linear equation for nonlinear dynamical system}
\setcounter{equation}{0}
\renewcommand{\theequation}{C\arabic{equation}}
For simplicity, we consider a finite-dimensional, but a general non-autonomous dynamical system. 
Let $\bm{z}(t)\in\mathbb{R}^{m}$ be real-valued variables. The nonlinear dynamical equation is then given by 
\begin{equation}
  d_t \bm{z}(t) = \bm{f}\bigl(t,\bm{z}(t)\bigr),
\end{equation}
where $\bm{f}$ is assumed to be continuously differentiable with respect to $\bm{z}$ in a domain of interest. 
The flow map associated with Eq. (C1) is denoted as $\Phi_{t,t_0}$: 
\begin{equation}
  \bm{z}(t) = \Phi_{t,t_0} \bm{z}(t_0).
\end{equation}

In order to derive the tangent, i.e., linearized, dynamics, we perturb the initial state as
$\bm{z}^{(\veps)}(t_0) := \bm{z}(t_0) + \veps\,\delta\bm{z}(t_0)$, 
and define the corresponding perturbed trajectory by 
$\bm{z}^{(\veps)}(t) = \Phi_{t,t_0}\bm{z}^{(\veps)}(t_0)$. 
We then introduce the variation along the reference trajectory as 
\begin{equation}
\delta\bm{z}(t)
:=
\left.d_\veps \bm{z}^{(\veps)}(t)\right|_{\veps=0}
=
\lim_{\veps\to 0}\veps^{-1}\left \{ \bm{z}^{(\veps)}(t)-\bm{z}(t) \right \}. 
\end{equation}
Differentiating Eq. (C1) with respect to $\veps$ and evaluating at $\veps=0$ yields the tangent linear equation: 
\begin{equation}
d_t \delta\bm{z}(t)
=
\bm{M}_f\bigl(t,\bm{z}(t)\bigr)\cdot\delta\bm{z}(t), 
\end{equation}
where $\bm{M}_f(t,\bm{z}):= \partial_{\bm{z}}\bm{f}(t,\bm{z})$ is the Jacobian of $\bm{f}$ with respect to $\bm{z}$. 
More generally, the corresponding object is the Fr\'echet derivative $D_{\bm{z}}\bm{f}(t,\bm{z})$ as a bounded linear map, 
and $\bm{M}_f$ is the matrix representation in finite dimensions. 

Connecting the tangent dynamics with the flow map, we define the tangent map $\bm{U}(t,t_0)$ along the reference
trajectory $\bm{z}(t)$ by
\begin{eqnarray}
  \delta\bm{z}(t) \! \! & = &  \! \! \bm{U}(t,t_0)\cdot\delta\bm{z}(t_0), \\
  \bm{U}(t,t_0)  \! \! & := &  \! \!
  \partial_{\bm{z}_0}\Phi_{t,t_0}\bigl(\bm{z}_0\bigr)\Big|_{\bm{z}_0=\bm{z}(t_0)},
\end{eqnarray}
where $\bm{U}(t_0,t_0)=\bm{I}_m$. 
By substituting Eq. (C5) into the tangent linear equation in Eq. (C4), we obtain an evolution equation for $\bm{U}(t,t_0)$: 
\begin{equation}
  d_t \bm{U}(t,t_0)
  = \bm{M}_f\bigl(t,\bm{z}(t)\bigr)\cdot\bm{U}(t,t_0). 
\end{equation}
A formal solution is then given by the time-ordered exponential,
\begin{equation}
  \bm{U}(t,t_0)
  =
  \mathcal{T}\exp\!\left [
    \int_{t_0}^{t} d\tau\,
    \bm{M}_f\bigl(\tau,\bm{z}(\tau)\bigr)
  \right ],
\end{equation}
where $\mathcal{T}$ denotes time ordering operator, i.e., $\mathcal{T}[f(t_2)g(t_0)h(t_1)] = f(t_2)h(t_1)g(t_0)$. 
Accordingly, the solution of $\delta \bm{z}(t)$ yields 
\begin{equation}
  \delta\bm{z}(t)
  =
  \mathcal{T}\exp\!\left [
    \int_{t_0}^{t} d\tau\,
    \bm{M}_f\bigl(\tau,\bm{z}(\tau)\bigr)
  \right ]\cdot \delta\bm{z}(t_0).
\end{equation}

If the system is autonomous, i.e., $\bm{f}(t,\bm{z}) = \bm{f}(\bm{z})$, the Jacobian also becomes $\bm{M}_f(t,\bm{z}) = \bm{M}_f(\bm{z})$
and the flow map depends only on the time difference $t-t_0$.
In this case, the tangent map still holds the similar representation
\begin{equation}
  \bm{U}(t,t_0)
  =
  \mathcal{T}\exp\!\left [
    \int_{t_0}^{t} d\tau\,
    \bm{M}_f\bigl(\bm{z}(\tau)\bigr)
  \right],
\end{equation}
and it reduces to an ordinary exponential only when $\bm{M}_f(\bm{z}(\tau))$ commutes with itself at different times along the trajectory.
In particular, for a linear autonomous system $d_t\bm{z}=\bm{K}\cdot\bm{z}$ with a constant matrix $\bm{K}$,
one obtains $\bm{U}(t,t_0)=\exp[(t-t_0)\bm{K}]$.

In addition, for an autonomous system, the growth rate of an infinitesimal perturbation can be quantified by Lyapunov exponents. 
The so-called maximal Lyapunov exponent is calculated by
\begin{equation}
  \lambda_{\max}
  :=
  \lim_{t\to\infty}\frac{1}{t-t_0}\ln \frac{\|\delta\bm{z}(t)\|}{\|\delta\bm{z}(t_0)\|}
  =
  \lim_{t\to\infty}\frac{1}{t-t_0}\ln \|\bm{U}(t,t_0)\|, 
\end{equation}
where $\|\cdot \|$ means a norm for vector or matrix.  
When we consider a canonical Hamiltonian dynamics with $\bm{z}=(\bm{q},\bm{p})\in\mathbb{R}^{2n}$, 
the dynamical equation for the Hamiltonian $H$ is given by 
\begin{eqnarray}
  d_t \bm{z}(t) \!\! & =\!\! & \bm{J}\cdot \partial_{\bm{z}} H\bigl(\bm{z}(t),t\bigr), \\
\bm{J} \!\! & = & \!\! 
\begin{pmatrix}
\bm{0} & \bm{I}_n\\
-\bm{I}_n & \bm{0}
\end{pmatrix}. 
\end{eqnarray}
The Jacobian appearing in the tangent linear equation is given by the Hessian:  
\begin{equation}
  \bm{M}_f(t,\bm{z})
  =
  \bm{J}\cdot \partial_{\bm{z}}\partial_{\bm{z}} H(\bm{z},t), 
\end{equation}
where the formal solution is the same as Eqs. (C8)--(C9). 
%
%
%
%
%
\section{D. Derivation of analytic expression for $C_{AB}^{\rm HM}(t,t_0)$}
\setcounter{equation}{0}
\renewcommand{\theequation}{D\arabic{equation}}
In Sec. 3.3, the time-lag dependence of the classical-limit OTOC is shown in terms of the quasilinear and strong zonal-flow shear dynamics.
In this appendix, we give an analytic derivation. 
It is shown that the OTOC grows at short time, and approaches a finite saturated value with an inverse-square algebraic dependence at later time. 

We start from the first-order zonal response in Eq. (47): 
\begin{equation}
\delta\bar q_B(t,x^{\prime})
=
- \! \int_{t_0}^{t}d\tau\,
P\left[
\{\tilde\phi,\delta\tilde q_B\}_{xy}
+
\{\delta\tilde\phi_B,\tilde q\}_{xy}
\right],
\end{equation}
where $x^{\prime}=x-x_0$ and $P$ is the zonal projection as the $y$-average.
The Fourier representation is given as $\delta\bar q_B(t,x^{\prime}) = \sum_{\bm{k}_{\rm zf}}\widehat{\delta\bar q}_{Bk_{\rm zf}}(t)\exp[ik_{\rm zf}x^{\prime}]$ 
with $\bm{k}_{\rm zf}:=(k_{\rm zf},0)$.   
In the local constant flow-shear approximation, the non-zonal fields, which correspond to Eqs. (42) and (43), are represented as 
\begin{eqnarray}
\tilde\phi(t,x^{\prime},y)
 \! \!  &=&  \! \! 
\sum_{\bm{k}^{(1)}}^{({\rm nz})}
\widehat{\tilde\phi}_{\bm{k}^{(1)}}(t)
\exp\!\left[i\left (k_x^{(1)}(t)x^{\prime}+k_y^{(1)}y \right )\right]
\exp\!\left[i k_y^{(1)}\!\int_{t_0}^{t}\! d\tau\, U_0(\tau)\right],
\\
\delta\tilde q_B(t,x^{\prime},y)
 \! \!  &=&  \! \! 
\sum_{\bm{k}^{(2)}}^{({\rm nz})}
\widehat{\delta\tilde q}_{B\bm{k}^{(2)}}(t)
\exp\!\left[i\left (k_x^{(2)}(t)x^{\prime}+k_y^{(2)}y\right )\right]
\exp\!\left[i k_y^{(2)}\! \int_{t_0}^{t}\! d\tau\, U_0(\tau)\right],
\\
k_x^{(\alpha)}(t)
\! \!  &:=&  \! 
k_{x0}^{(\alpha)}+k_y^{(\alpha)}S_0(t-t_0),
\qquad
\alpha=1,2. 
\end{eqnarray}
The independent perpendicular wavenumber labels in the Fourier coefficients for the unperturbed($\alpha\! =\! 1$) and perturbed($\alpha\! =\! 2$) fields 
are denoted with $\bm{k}^{(\alpha)}:=\left(k_{x0}^{(\alpha)},k_y^{(\alpha)}\right)$, 
and 
$\sum_{\bm{k}^{(\alpha)}}^{({\rm nz})}:=
\sum_{k_{x0}^{(\alpha)}}\sum_{k_y^{(\alpha)}\neq0}$ means the summation for non-zonal wavenumber modes. 
Substituting Eqs. (D2)--(D4) into the first Poisson bracket in Eq. (D1), a straightforward algebra leads to  
\begin{eqnarray}
P\{\tilde\phi,\delta\tilde q_B\}_{xy}
\! \! &=&  \! \!
\sum_{\bm{k}_{\rm zf}}
\sum_{\bm{k}^{(1)}}^{({\rm nz})}
\sum_{\bm{k}^{(2)}}^{({\rm nz})}
\delta_{\bm{k}^{(1)}+\bm{k}^{(2)}-\bm{k}_{\rm zf},\bm{0}}
\left [ \frac{k_{x0}^{(2)}k_{y}^{(1)}-k_{x0}^{(1)}k_{y}^{(2)}}{1+\{k_{x}^{(1)}(t)\}^2 + \{k_{y}^{(1)}\}^2} \right ]\,
\widehat{\tilde q}_{\bm{k}^{(1)}}(t)\,
\widehat{\delta \tilde q}_{B\bm{k}^{(2)}}(t) \nonumber \\
 \! \! &\times&  \! \!
\exp \left [ i\left \{ k_{x0}^{(1)}+k_{x0}^{(2)} + \left (k_{y}^{(1)}+k_{y}^{(2)} \right ) S_0(t-t_0) \right \}x^{\prime} +i(k_{y}^{(1)}+k_{y}^{(2)})y \right ] \nonumber \\
 \! \! &\times&  \! \!
\exp \left [ i\left (k_{y}^{(1)}+k_{y}^{(2)} \right )\! \int_{t_0}^{t} \! d\tau\, U_0(\tau) \right ] \nonumber \\
 \! \! &=&  \! \!
\sum_{\bm{k}_{\rm zf}}
\sum_{\bm{k}^{(1)}}^{({\rm nz})}
\left [ \frac{k_{\rm zf}k_{y}^{(1)}}{1+\{k_{x}^{(1)}(t)\}^2 + \{k_{y}^{(1)}\}^2} \right ]\,
\widehat{\tilde q}_{\bm{k}^{(1)}}(t)\,
\widehat{\delta \tilde q}_{B\bm{k}_{\rm zf} - \bm{k}^{(1)}}(t)
\exp[ik_{\rm zf}x^{\prime}], 
\end{eqnarray}
where Eq. (19) [or equivalently Eq. (44)] is used to replace $\tilde \phi$ by $\tilde q$. 
The second Poisson bracket in Eq. (D1) is calculated in the same manner, 
and one finds the Fourier coefficient of the zonal response $\widehat{\delta\bar q}_{Bk_{\rm zf}}(t)$ as follows: 
\begin{eqnarray}
\widehat{\delta\bar q}_{Bk_{\rm zf}}(t)
 \! \! &=&  \! \!
-\!\int_{t_0}^{t}\!d\tau\,
\widehat \jj_{k_{\rm zf}}(\tau), \\
\widehat \jj_{k_{\rm zf}}(\tau)
 \! \! &=& \! \!
\sum_{\bm{k}}^{({\rm nz})}
 k_{\rm zf}k_y\,
\widehat{\tilde q}_{\bm{k}}(\tau)\,
\widehat{\delta\tilde q}_{B\bm{k}_{\rm zf}-\bm{k}}(\tau)
\left[
D_{\bm{k}}^{-1}(\tau)-D_{\bm{k}_{\rm zf}-\bm{k}}^{-1}(\tau)
\right], 
\end{eqnarray}
where 
\begin{eqnarray}
D_{\bm{k}}(\tau)
 \! \! &:=& \! \!
1+k_y^2+ \left[k_{x0}+k_yS_0(\tau-t_0)\right]^2, \nonumber \\
D_{\bm{k}_{\rm zf}-\bm{k}}(\tau)
\! \! &:=& \! \!
1+k_y^2+ \left[ k_{\rm zf}-k_{x0}-k_yS_0(\tau-t_0) \right]^2.
\end{eqnarray}
Note here that the single non-zonal wavenumber label is simply denoted by $\bm{k}=(k_{x0}, k_y)$. 

Considering $\mathcal P_{\bm{k}}(t):= \widehat{\tilde q}_{\bm{k}}(t)\, \widehat{\delta\tilde q}_{B\bm{k}_{\rm zf}-\bm{k}}(t)$ and Eq. (45), we obtain 
\begin{equation}
d_t\mathcal P_{\bm{k}}(t) =
ik_yG_0 \left[ D_{\bm{k}}^{-1}(t)-D_{\bm{k}_{\rm zf}-\bm{k}}^{-1}(t) \right] \mathcal P_{\bm{k}}(t),
\end{equation}
and the exact solution is given as 
\begin{eqnarray}
\mathcal P_{\bm{k}}(t)
\! \! &=& \! \!
\mathcal P_{\bm{k}}(t_0) \exp\!\left[ i k_yG_0\mathcal I_{\bm{k}}(t,t_0) \right], \nonumber \\
\mathcal I_{\bm{k}}(t,t_0) 
\! \! &:=&  \! \!
\int_{t_0}^{t}d\tau \left[ D_{\bm{k}}^{-1}(\tau)-D_{\bm{k}_{\rm zf}-\bm{k}}^{-1}(\tau) \right].
\end{eqnarray}
Then, using Eqs. (D6)--(D10), the zonal response is rewritten as 
\begin{eqnarray}
\widehat{\delta\bar q}_{Bk_{\rm zf}}(t)
\! \! &=&  \! \!
-\!\int_{t_0}^{t}\!d\tau\,
\widehat \jj_{k_{\rm zf}}(\tau) \nonumber \\
 \! \! &=&  \! \!
-\!\int_{t_0}^{t}\!d\tau\,
\left\{ \frac{k_{\rm zf}}{iG_0}\sum_{\bm{k}}^{({\rm nz})}d_{\tau}\mathcal P_{\bm{k}}(\tau) \right \} \nonumber \\
 \! \! &=&  \! \!
\frac{i k_{\rm zf}}{G_0} \sum_{\bm{k}}^{({\rm nz})} \left \{ \mathcal P_{\bm{k}}(t)-\mathcal P_{\bm{k}}(t_0) \right \} \nonumber \\
\! \! &=&  \! \!
\frac{i k_{\rm zf}}{G_0} \sum_{\bm{k}}^{({\rm nz})} \mathcal P_{\bm{k}}(t_0) \left \{ \exp\!\left[i k_yG_0\mathcal I_{\bm{k}}(t,t_0)\right]-1 \right \}.
\end{eqnarray}
The time-dependent phase factor $\mathcal{I}_{\bm{k}}(t,t_0)$ is analytically integrated as 
\begin{eqnarray}
\mathcal I_{\bm{k}}(t,t_0)
 \! &=&  \!
\frac{1}{k_yS_0K} \Bigg[ \tan^{-1}\!\left( \frac{k_{x0}+k_yS_0(t-t_0)}{K} \right) - \tan^{-1}\!\left( \frac{k_{x0}}{K} \right) \nonumber \\
&&
- \tan^{-1}\!\left( \frac{k_{\rm zf}-k_{x0}}{K} \right) + \tan^{-1}\!\left( \frac{k_{\rm zf}-k_{x0}-k_yS_0(t-t_0)}{K} \right) \Bigg],
\end{eqnarray}
where $K:=(1+k_{y}^2)^{\frac{1}{2}}$. 
The limit value for $t\to \infty$ is 
\begin{equation}
\mathcal I_{\bm{k}}(\infty) := \lim_{t \to \infty} \mathcal I_{\bm{k}}(t,t_0)
= -\frac{1}{k_yS_0K} 
\left[ \tan^{-1}\!\left( \frac{k_{x0}}{K} \right) + \tan^{-1}\!\left( \frac{k_{\rm zf}-k_{x0}}{K} \right) \right].
\end{equation}
It is noted that the derivation provided here involves a full Fourier convolution in the two Poisson brackets in Eq. (D1), 
covering a solution under the single-mode approximation in earlier work\cite{Mikhailenko}. 

We now relate $\widehat{\delta\bar q}_{Bk_{\rm zf}}(t)$ to the zonal-flow-energy response $\delta E_{\rm ZF}(t\,|\,t_0,Q_{\rm NZ})$. 
In the form of Fourier expansion with a domain size $L_xL_y$, Eq. (38) is rewritten as  
\begin{equation}
\delta E_{\rm ZF}(t\,|\,t_0,Q_{\rm NZ}) = L_xL_y \sum_{k_{\rm zf}} \widehat U_{-k_{\rm zf}}(t)\, \widehat{\delta U}_{k_{\rm zf}}(t), 
\end{equation}
where $U(t,x^{\prime})= \sum_{k_{\rm zf}} \widehat U_{k_{\rm zf}}(t)\exp[i k_{\rm zf}x^{\prime}]$ are used.
Applying $\widehat{\delta U}_{k_{\rm zf}}(t) = - i k_{\rm zf}(1+k_{\rm zf}^2)^{-1} \widehat{\delta\bar q}_{Bk_{\rm zf}}(t)$ with Eq. (D11), one finds 
\begin{equation}
\delta E_{\rm ZF}(t\,|\,t_0,Q_{\rm NZ})
=
L_xL_y \sum_{k_{\rm zf}} \frac{k_{\rm zf}^2}{(1+k_{\rm zf}^2)G_0} \widehat U_{-k_{\rm zf}}(t) \sum_{\bm{k}}^{({\rm nz})}
\mathcal P_{\bm{k}}(t_0)
\left\{
\exp\!\left[i k_yG_0\mathcal I_{\bm{k}}(t,t_0)\right]-1
\right \}.
\end{equation}
Thus, the classical-limit OTOC for H-M turbulence is expressed as  
\begin{equation}
C_{AB}^{\rm HM}(t,t_0)
=
\left\langle
\left[
\delta E_{\rm ZF}(t\,|\,t_0,Q_{\rm NZ})
\right]^2
\right\rangle_{\rm ens}.
\end{equation}
It should be noted that Eqs. (D15) and (D16) are still exact within the local quasilinear ansatz. 
In the following asymptotic estimates of the short-time or long-time limit, 
we additionally assume the time scale of amplitude and the strength of shearing for background zonal flows.   

Assuming that the temporal variation of the background zonal-flow amplitude is much slower than that in the non-zonal perturbations, i.e., 
$\widehat U_{-k_{\rm zf}}(t)\simeq \widehat U_{-k_{\rm zf}}(t_0)$, 
we can evaluate two asymptotics for time-lag dependence in the classical-limit OTOC, where $\Delta t := t-t_0 \geqslant 0$. 
First, we consider the short-time limit in $C_{AB}^{\rm HM}(t,t_0)$. 
Expanding the integrand in Eq. (D12) gives 
\begin{equation}
\mathcal I_{\bm{k}}(t,t_0)
=\left[
\frac{1}{D_{0\bm{k}}}-
\frac{1}{D_{0\bm{k}_{\rm zf}-\bm{k}}}
\right]
\Delta t
-
 k_yS_0
\left[
\frac{k_{x0}}{D_{0\bm{k}}^2}
+
\frac{k_{\rm zf}-k_{x0}}{D_{0\bm{k}_{\rm zf}-\bm{k}}^2}
\right]
(\Delta t)^2
+
O((\Delta t)^3), 
\end{equation}
where $D_{0\bm{k}}:=1+k_y^2+k_{x0}^2$ and $D_{0\bm{k}_{\rm zf}-\bm{k}}:=1+k_y^2+(k_{\rm zf}-k_{x0})^2$. 
Using $\exp[i k_yG_0\mathcal I_{\bm{k}}]-1 = i k_yG_0\mathcal I_{\bm{k}}+O((\Delta t)^2)$, we obtain 
\begin{equation}
C_{AB}^{\rm HM}(t,t_0) \simeq \left\langle \mathcal A(t_0)^2\right\rangle_{\rm ens} (\Delta t)^2 +O((\Delta t)^3), \quad \Delta t \ll 1, 
\end{equation}
where 
\begin{equation}
\mathcal A(t_0) := i L_xL_y \sum_{k_{\rm zf}} \frac{k_{\rm zf}^2}{1+k_{\rm zf}^2} \widehat U_{-k_{\rm zf}}(t_0) \sum_{\bm{k}}^{({\rm nz})} k_y\mathcal P_{\bm{k}}(t_0)
\left[ D_{0\bm{k}}^{-1}-D_{0\bm{k}_{\rm zf}-\bm{k}}^{-1} \right].
\end{equation}
Thus, we find that the OTOC grows quadratically with respect to $\Delta t$ in the short-time asymptotic expansion.  
If the leading coefficient $\mathcal A(t_0)$ vanishes owing to wavenumber symmetry or ensemble averaging, the next order controls the faster initial growth. 

Next, we consider the long-time limit of $\Delta t \gg 1$. 
By using the large-argument expansion formula of $\tan^{-1}z=(\pi/2){\rm sgn}(z)-z^{-1}+(1/3)z^{-3}+O(z^{-5})$ with $|k_yS_0\Delta t| \gg 1$, we find 
\begin{equation}
\mathcal I_{\bm{k}}(t,t_0)-\mathcal I_{\bm{k}}(\infty)
=
\frac{k_{\rm zf}}{k_y^3S_0^3(\Delta t)^2}
-
\frac{k_{\rm zf}(2k_{x0}-k_{\rm zf})}{k_y^4S_0^4(\Delta t)^3}
+
O((\Delta t)^{-4}).
\end{equation}
Since $\delta E_{\rm ZF}(t\,|\,t_0,Q_{\rm NZ})$ saturates in time due to $\lim_{t \to \infty}\mathcal{I}_{\bm{k}}(t,t_0) = \mathcal{I}_{\bm{k}}(\infty)$, 
the difference between the saturated response and the finite-time response is, therefore, expressed as 
\begin{eqnarray}
&&
\delta E_{\rm ZF}(\infty\,|\,t_0,Q_{\rm NZ})
-
\delta E_{\rm ZF}(t\,|\,t_0,Q_{\rm NZ})
\nonumber\\
&&\simeq
L_xL_y
\sum_{k_{\rm zf}}
\frac{k_{\rm zf}^2}{(1+k_{\rm zf}^2)G_0}
\widehat U_{-k_{\rm zf}}(t_0)
\sum_{\bm{k}}^{({\rm nz})}
\mathcal P_{\bm{k}}(t_0)
\nonumber\\
&&\quad\times
\left \{
\exp\!\left[i k_yG_0\mathcal I_{\bm{k}}(\infty)\right]
-
\exp\!\left[i k_yG_0\mathcal I_{\bm{k}}(t,t_0)\right]
\right \} \nonumber \\
&&=
L_xL_y
\sum_{k_{\rm zf}}
\frac{k_{\rm zf}^2}{(1+k_{\rm zf}^2)G_0}
\widehat U_{-k_{\rm zf}}(t_0)
\sum_{\bm{k}}^{({\rm nz})}
\mathcal P_{\bm{k}}(t_0)
\nonumber\\
&&\quad\times
\exp\!\left[i k_yG_0\mathcal I_{\bm{k}}(\infty)\right]
\left \{
1-
\exp\!\left[i k_yG_0\Delta\mathcal I_{\bm{k}}(t,t_0)\right]
\right \}, 
\end{eqnarray}
where $\Delta\mathcal I_{\bm{k}}(t,t_0) := \mathcal I_{\bm{k}}(t,t_0)-\mathcal I_{\bm{k}}(\infty)$.  
Expanding as $1- \exp[i k_yG_0\Delta\mathcal I_{\bm{k}}] = -i k_yG_0\Delta\mathcal I_{\bm{k}} +O((\Delta\mathcal I_{\bm{k}})^2)$, 
we obtain 
\begin{equation}
\delta E_{\rm ZF}(\infty\,|\,t_0,Q_{\rm NZ})
-
\delta E_{\rm ZF}(t\,|\,t_0,Q_{\rm NZ})
=
O\left[S_0^{-3}(\Delta t)^{-2}\right], \quad \Delta t \gg 1.
\end{equation}
Subsequently, one can consider 
\begin{eqnarray}
C_{AB}^{\rm HM}(\infty,t_0)-C_{AB}^{\rm HM}(t,t_0)
\! \! &=&  \! \!
\left\langle \left[ \delta E_{\rm ZF}(\infty\,|\,t_0,Q_{\rm NZ}) \right]^2 \right\rangle_{\rm ens}
-\left\langle \left[ \delta E_{\rm ZF}(t\,|\,t_0,Q_{\rm NZ}) \right]^2 \right\rangle_{\rm ens}
\nonumber\\
\! \! &=&  \! \!
\left\langle
\left[
\delta E_{\rm ZF}(\infty\,|\,t_0,Q_{\rm NZ})
-
\delta E_{\rm ZF}(t\,|\,t_0,Q_{\rm NZ})
\right]
\right.
\nonumber\\
&&\qquad\left.
\times
\left[
\delta E_{\rm ZF}(\infty\,|\,t_0,Q_{\rm NZ})
+
\delta E_{\rm ZF}(t\,|\,t_0,Q_{\rm NZ})
\right]
\right\rangle_{\rm ens}, 
\end{eqnarray}
where $C_{AB}^{\rm HM}(\infty,t_0):=\lim_{t \to \infty} C_{AB}^{\rm HM}(t,t_0)$. 
It is obvious from Eq. (D13) that $\mathcal{I}_{\bm{k}}(\infty) = O(S_0^{-1})$, and then the strong-shear approximation of $|G_0/S_0| \ll 1$ yields 
$\exp[i k_yG_0\mathcal I_{\bm{k}}(\infty)]-1 = i k_yG_0\mathcal I_{\bm{k}}(\infty)+O(S_0^{-2})$.  
By using these relations and Eqs. (D15) and (D22), we can evaluate as $\delta E_{\rm ZF}(\infty\,|\,t_0,Q_{\rm NZ}) = O(S_0^{-1})$ and therefore
\begin{equation}
\delta E_{\rm ZF}(\infty\,|\,t_0,Q_{\rm NZ}) + \delta E_{\rm ZF}(t\,|\,t_0,Q_{\rm NZ}) =O(S_0^{-1}). 
\end{equation}
Combining Eqs. (D22)--(D24), the long-time limit of the OTOC is finally obtained as 
\begin{equation}
C_{AB}^{\rm HM}(t,t_0) = C_{AB}^{\rm HM}(\infty,t_0) + O\left[S_0^{-4}(\Delta t)^{-2}\right], \quad \Delta t \gg 1.
\end{equation}
It is emphasized that the OTOC asymptotically approaches a finite saturated value with an inverse-square algebraic dependence.
The physical origin is that the zonal-flow shear transfers the non-zonal perturbation to higher radial wavenumbers, 
thereby suppressing the remaining non-zonal-to-zonal feedback.
%
%

\section*{ACKNOWLEDGMENTS}
\vspace{-0.4cm}
The author thanks Dr. Go Yatomi, Dr. Kota Yanagihara, and Dr. Shun Arai for fruitful discussions on Weyl-Wigner-Moyal formalism. 
This work is supported by the MEXT Japan, Grant No. 21H04458 
and by the NIFS collaborative Research Programs (NIFS24KIPM005, NIFS26KISI005), 
and by JST (Moonshot R\&D Program) Japan, Grant Number JPMJMS24A1.


\end{document}